\documentclass[aps, superscriptaddress, twocolumn, pre]{revtex4-1}
\usepackage{amsmath,graphicx,color,epsfig,latexsym,bm,ulem,dutchcal,subfigure,float,tikz,eucal, mathpazo,times, braket}
\usepackage[colorlinks=true, linkcolor = blue, urlcolor  = blue, citecolor = blue, anchorcolor = red]{hyperref}   
\newcommand{\abs}[1]{\left\vert#1\right\vert}

\usetikzlibrary{positioning}

\begin{document}
	
	
	\title{Heat rectification by two qubits coupled with Dzyaloshinskii--Moriya interaction}
	\author{Vipul Upadhyay}
	\affiliation{Department of Physics, Indian Institute of Technology Delhi, Hauz Khas 110 016, INDIA}
	\author{M. Tahir Naseem} 
	\affiliation{Department of Physics, Ko\c{c} University, 34450 Sariyer, Istanbul, Turkey}
	\author{Rahul Marathe}\email{maratherahul@physics.iitd.ac.in} \affiliation{Department of Physics, Indian Institute of Technology Delhi, Hauz Khas 110 016, INDIA} 
	\author{\"Ozg\"ur E. M\"ustecapl\ifmmode \imath \else \i \fi{}o\ifmmode \breve{g}\else \u{g}\fi{}lu}
	\email{omustecap@ku.edu.tr}
	\affiliation{Department of Physics, Ko\c{c} University, 34450 Sariyer, Istanbul, Turkey}
	
	\begin{abstract}{We investigate heat rectification in a two-qubit system coupled via the Dzyaloshinskii-Moriya (DM) interaction. We derive analytical expressions for heat currents and thermal rectification and provide possible physical mechanisms behind the observed results. We show that the anisotropy of DM interaction in itself is insufficient for heat rectification, and some other form of asymmetry is needed. We employ off-resonant qubits as the source of this asymmetry. We find the regime of parameters for higher rectification factors by examining the analytical expressions of rectification obtained from a global master equation solution. In addition, it is shown that the direction and quality of rectification can be controlled via various system parameters. Furthermore, we compare the influence of different orientations of the DM field anisotropy on the performance of heat rectification. Finally, we investigate the possible interplay between quantum correlations and the performance of the quantum thermal rectifier. We find that asymmetry in the coherences is a fundamental resource for the performance of the quantum thermal rectifier.}
	\end{abstract}
	
	\maketitle
	
	\section{Introduction}\label{sec:model}
	
Manipulation of heat at the nanoscale, particularly thermal rectification by heat diodes, is currently a subject of intense theoretical~\cite{PhysRevLett.88.094302, classical1, PhysRevLett.93.184301, spin_boson_thermal_rectifier, PhysRevB.74.214305, Scheibner_2008, PhysRevLett.100.105901, PhysRevE.80.041103, PhysRevB.79.144306, PhysRevLett.102.095503, PhysRevB.80.172301, PhysRevB.81.205321, PhysRevB.74.214305, PhysRevLett.104.154301, PhysRevLett.107.173902, RevModPhys.84.1045, PhysRevB.88.094427, doi:10.1063/1.4817258, Zhang_2013, Thingna_2013, XXZ2, PhysRevE.89.062109, PhysRevB.92.045309, PhysRevE.94.042135, PhysRevE.94.042122, Newexp_1,lindblad_equation, PhysRevE.96.012114, PhysRevB.98.035414, PhysRevLett.120.200603, Motz_2018,lab,Newt_3, segmentedchain, Muhammad, PhysRevB.99.035129, PhysRevE.99.032126, PhysRevE.99.032116, PhysRevResearch.2.033285, PhysRevE.102.062146, PhysRevE.101.062122, PhysRevB.103.104304, PhysRevE.103.052130, PhysRevB.103.155434, PhysRevApplied.15.054050, PhysRevE.103.012134, PhysRevB.103.115413, tupkary2021} and experimental~\cite{experiment4, JWJing_experiment, Chen2014, Martinez15, Seif2018, Newexp_3, Senior2020, Maillet2020} research. The theoretical studies explain the change of heat current direction and magnitude when the thermal bias is reversed due to inherent asymmetry and non-linearity  in the physical models ~\cite{RevModPhys.84.1045, dhar}. The most common naturally occurring interaction used in such quantum models is the Heisenberg exchange interaction~\cite{segmentedchain,lindblad_equation,XXZ2} where the asymmetry stems from either the different  on-site magnetic fields~\cite{lindblad_equation} or from the different coupling constants of the subsystem with their respective baths~\cite{spin_boson_thermal_rectifier}. 
	
	Recently, an artificially designed system containing inherent asymmetry in the Hamiltonian itself has also been proposed to operate under symmetric system-bath couplings and resonant (identical) subsystems, which allows for the rectification of large heat currents~\cite{Muhammad}. The model is based upon coupling $z$-component of a spin to $x$-component of the other spin in a two-qubit system. The coupling is, therefore, asymmetric under the exchange of the spins. This artificial model takes only one of the two terms of the $y$ component of the two spins' cross product. Intriguingly, there is a naturally existing, entirely physical interaction depending on the cross product of the spins, known as Dzyaloshinskii Moriya (DM) interaction~\cite{dzyaloshinsky_thermodynamic_1958, moriya_anisotropic_1960}. It is antisymmetric under the exchange of spins due to its cross-product dependence. Accordingly, we ask whether DM interaction's anti-symmetry is sufficient per se to generate heat rectification and, even if it is not, how it influences the heat conduction properties. To answer this question, we take a simple system of two DM-coupled spin-$1/2$ particles (qubits) with the DM field along $z$ direction and analytically derive the quantum master equation and heat current expression for our model. We also derive an analytical expression for the {rectification} factor and discuss the behavior of our model under various parameter limits, providing possible physical mechanisms for the same. We provide the possible configuration for optimizing the working of our {\it heat rectifier}. Finally, we explore the role of change of anisotropy field direction on the heat current and rectification ability. 
	
	Some effects of DM interaction on heat transport in spin chains have been studied \cite{Hui_Ping_2006,Li2012}. In particular, possible control of heat rectification using DM interaction in a system of two quantum dots has been proposed \cite{CHEN201558}. Another study of magnetic thermal rectification in a single molecule magnet concluded that thermal rectification is possible due to anti-symmetry of DM interaction \cite{XU2016107}. On the other hand, these studies include the exchange interaction next to the DM interaction, as such an interaction is naturally occurring. However, this may not reveal the role of the DM interaction on heat current per se. Here, we present a more systematic analysis by focusing on the DM interaction alone and investigate symmetry in heat flow in various parameter regimes. In addition, our analysis based on the derivation of a global master equation \cite{Levy_2014,PhysRevA.98.052123} and analytical results for both heat currents and rectification reveals the underlying mechanism that leads to thermal rectification. Based on the analytical results of rectification, we find the parameters regime for higher rectification factors, thereby guiding further development of thermal devices based on DM interaction. We believe such systematic analysis has not been done in previous studies of DM interaction-based heat rectifiers. Such a methodological examination of the DM interaction can also be experimentally feasible ~\cite{NatureDMexperiment}. Finally, we study the possible effects of the stationary quantum correlations between the qubits on the performance of the thermal rectifier. We find coherences are asymmetrical under the change of temperature bias, and this asymmetry is sufficient for the emergence of thermal rectification.

	\begin{figure}[!htbp]
		\begin{center}
			\leavevmode
			\includegraphics[width=0.5\textwidth,angle=0]{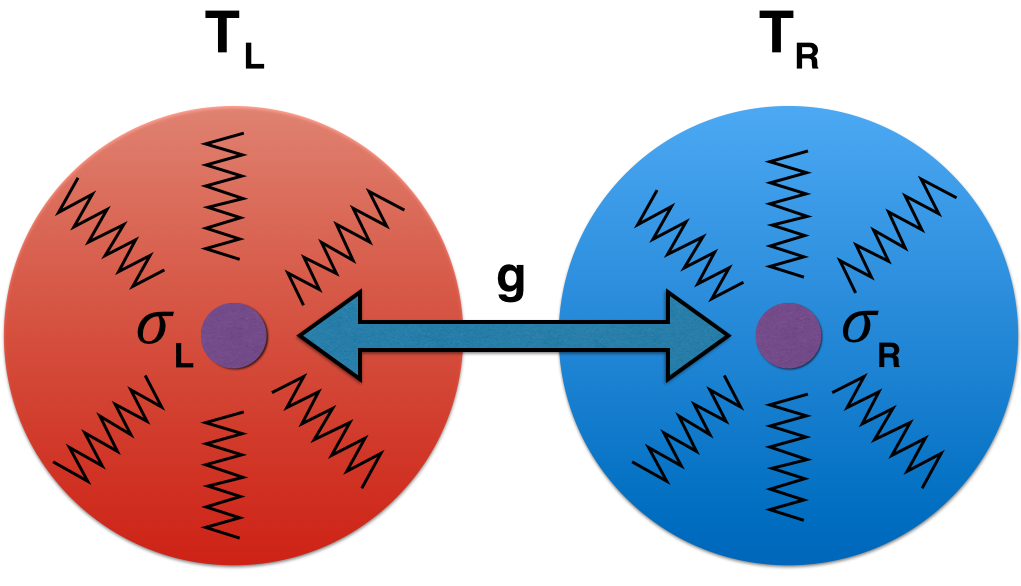}
			\caption{\label{model}(Color online) A schematic of the model consisting of two qubits coupled by the Dzyaloshinskii Moriya (DM) interaction with on-site (local) magnetic fields. Each qubit is attached to its own local bath.}
		\end{center}
	\end{figure}

The rest of the paper is organized as follows. {In Sec.~\ref{sec::model},} we introduce our model and physical system. Section~\ref{sec:MasterEq} gives the derivation of the master equation describing the open system dynamics of the model. In Sec.~ \ref{section: Results}, we discuss the analytical solution of the master equation in a series of subsections. First, Sec.~\ref{subsection::Current Results} presents the heat currents, followed by the rectification abilities of the model system in Sec.~\ref{subsection::Rectification Results}. Sec.~\ref{subsection::Effect of anisotropy} compares the models with anisotropy fields in different directions. In Sec.~\ref{section::coherences}, we investigate the possible role of the quantum correlations on heat rectification. We conclude in Sec.~\ref{sec:conclusion}. Additional details of the derivation of the master equation and the heat currents are provided in  Appendices~\ref{section:Appendix A} and~\ref{App:Appendix:B}, respectively.

	\section{The Model}\label{sec::model}

Our physical system consists of two spin-$1/2$ particles (qubits) coupled via the Dzyaloshinskii-Moriya (DM) interaction in the presence of on-site magnetic fields, which is effectively realizable in {nuclear magnetic resonance} (NMR) experiments~\cite{NatureDMexperiment}. Each qubit interacts with its {own} (local) bath at different temperatures, as illustrated in Fig.~\ref{model}. The Hamiltonian of the system is expressed as 
	
	\begin{eqnarray}\label{system}
	H_S=H_0+H_\text{DM},
	\end{eqnarray}
	where $H_0$ is the Hamiltonian of the non-interacting qubits, ({we} take reduced Planck constant as $\hbar=1$)
	\begin{eqnarray}
	\hat{H}_0&=&\frac{\omega_L}{2} \hat{\sigma}_L^z\otimes \hat{I}_R+\hat{I}_L\otimes\frac{\omega_R}{2} \hat{\sigma}_R^z,
	\end{eqnarray}
	and the DM interaction is described by
	the Hamiltonian,
	\begin{eqnarray}\label{eq:DMmodel}
	H_\text{DM}= g(\hat{\sigma}_L^x \hat{\sigma}_R^y-\hat{\sigma}_L^y\hat{\sigma}_R^x).
	\end{eqnarray}
	Here, $\hat{\sigma}_i^\alpha$'s denote the $\alpha=x,y,z$ components of the Pauli spin-$1/2$ operators for
	the left ($i=L$) and right ($i=R$) qubits. The unit operators are denoted by $\hat{I}_i$. We assume on-site magnetic fields can be used for locally distinct frequencies $\omega_L$ and $\omega_R$ for the left and right qubits, respectively. The DM coupling coefficient $g$ corresponds to the case where the DM anisotropy field is aligned in the $z$-direction so that the general DM interaction $\bm{D}\cdot (\bm{\sigma}_L\times\bm{\sigma}_R)$
	reduces to Eq.~(\ref{eq:DMmodel}) with $\bm{D}=g\bm{\hat{z}}$. 
	\begin{figure}[!htbp]
		\begin{center}
			\leavevmode
			\includegraphics[width=0.5\textwidth,angle=0]{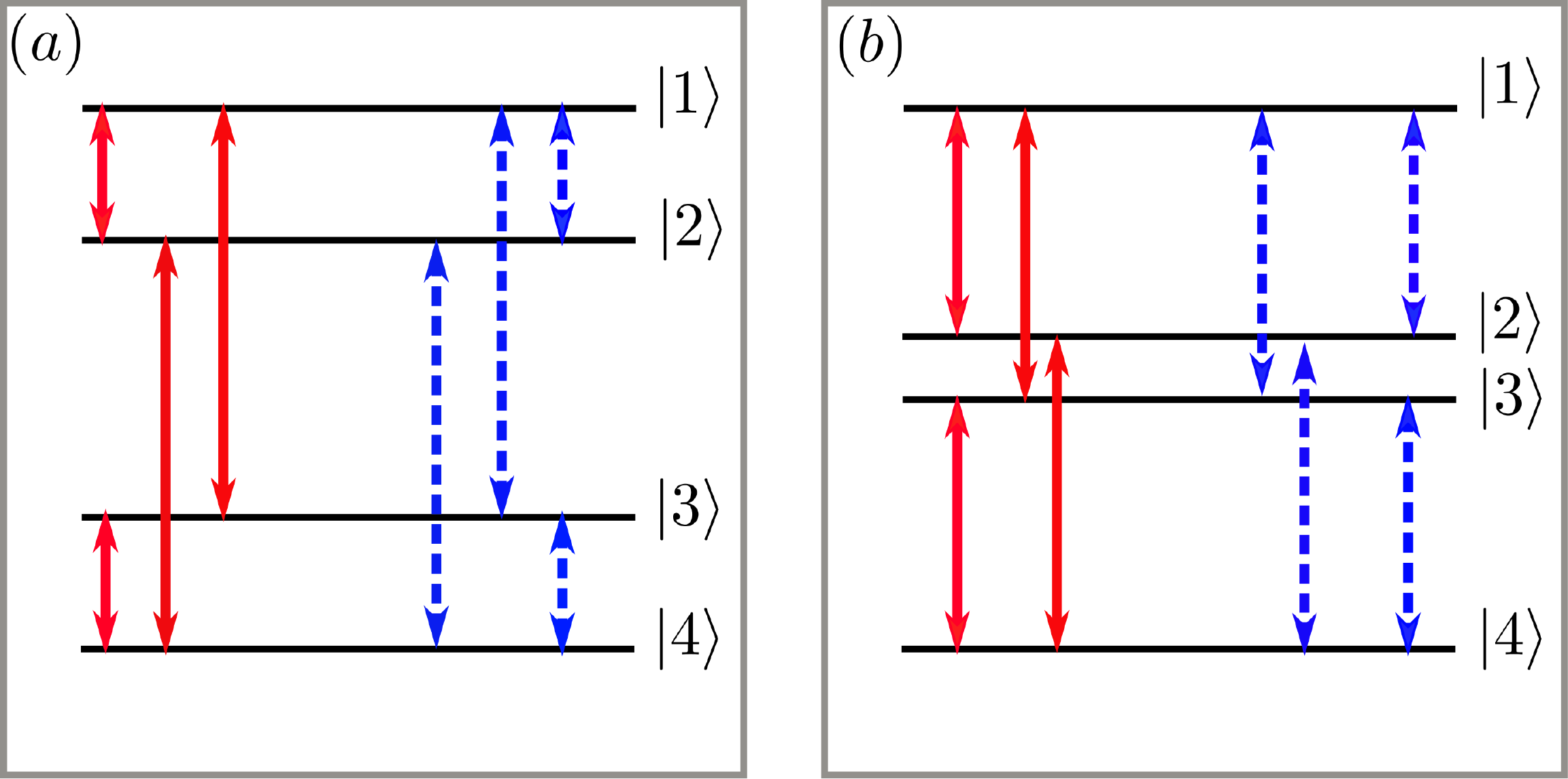}
			\caption{\label{fig:engylev}(Color online) Energy transitions induced by left (solid lines) and right (dashed lines) baths for weakly coupled (a)  off-resonant qubits, and (b) for nearly-resonant qubits.}
		\end{center}
	\end{figure}
	
	We take the 
	{spin-boson} model to describe dissipative coupling of each qubit to its respective local bath. Free Hamiltonian of each bath $B_i$ 
	{is given by}
	\begin{equation}
	\hat{H}_{B_i}=\sum_{n} \omega _n \hat{a}_{i,n}^{\dagger}\hat{a}_{i,n},
	\end{equation}  
	where {$\hat{a}_{i,n}^{\dagger}\text{ }(\hat{a}_{i,n})$} are the bosonic creation (annihilation) operators of the $n$-th mode of the $i^{th}$ bath. 
	The {system-bath} interaction is described by
	\begin{equation}
	\hat{H}_{SB}=\sum_{i,n} g_{i,n}\hat{\sigma}_x^i\otimes (\hat{a}^{\dagger}_{i,n}+\hat{a}_{i,n}),
	\end{equation}
	where $g_{i,n}$ represents the coupling coefficient of the $n$-th mode of
	the bath $B_i$ to the qubit labeled with $i$. We assume symmetric couplings between the baths and the qubits such that $g_{L,n}=g_{R,n}\equiv g_{n}$.
	
	\section{Master Equation}
	\label{sec:MasterEq}
	
		In this section, we outline the derivation of master equation for our model. 
		The eigenvalues of the system Hamiltonian~(\ref{system}) are given by
		\begin{eqnarray}\label{eigenvalues}
		\pm\omega_S&:=&\pm\frac{\omega_L+\omega_R}{2},\nonumber\\ 
		\pm\Omega &:=&\pm\sqrt{4 g^2+ \omega_D^2},
		\end{eqnarray}
		where $\omega_D:=(\omega_L-\omega_R)/2$ is introduced for brevity of notations.
		The sign of $\omega_D$ tells us which qubit is at higher frequency.
		The  eigenvectors in computational basis associated with the eigenvalues are expressed as
		\begin{align}
		&|1\rangle:=|\omega_S\rangle=|++\rangle, \nonumber\\
		&|2\rangle:=|\Omega\rangle=\cos{\theta}|+-\rangle+i \sin{\theta}|-+\rangle, \nonumber\\
		&|3\rangle:=|-\Omega\rangle=i \sin{\theta}|+-\rangle+\cos{\theta}|-+\rangle, \nonumber\\
		&|4\rangle:=|-\omega_S\rangle=|--\rangle,
		\end{align}
		where the parameter $\theta$ is defined as
		\begin{align}\label{sine_cosine_define}
		\cos{\theta}=\frac{2g}{\sqrt{4g^2+(\omega_D-\Omega)^2}},\text{    }\sin{\theta}=\frac{\omega_D-\Omega}{\sqrt{4g^2+(\omega_D-\Omega)^2}}.
		\end{align}
		Energy transitions induced by left and right baths are presented in Fig.~\ref{fig:engylev}, which shows all possible transitions that can be induced by the thermal baths.

		The master equation for our model is derived under the {usual Born-Markov and secular approximations,} and can be written in the interaction picture as~\cite{breuer2002} (see Appendix~\ref{section:Appendix A}) 
		
		\begin{equation}\label{local}
		\frac{d}{dt}\hat{\rho}(t)=\mathcal{L}_{L}\hat{\rho}(t)+\mathcal{L}_{R}\hat{\rho}(t),
		\end{equation}
		where the Liouvillian superoperators are given by
		\begin{align}\label{Master}
		\mathcal{L}_{L}\hat{\rho}(t)&=
		\cos{^2\theta}(G_L(\omega_+) \mathcal{D}(\tilde{\sigma}_L^-)+G_L(-\omega_+) \mathcal{D}(\tilde{\sigma}_L^+))\nonumber\\&+\sin{^2\theta}(G_L(\omega_-)  \mathcal{D}(\tilde{\sigma}_L^z\tilde{\sigma}_R^-)+G_L(-\omega_-)  \mathcal{D}(\tilde{\sigma}_L^z\tilde{\sigma}_R^+)), \nonumber \\ 
		\mathcal{L}_{R}\hat{\rho}(t)&=\cos{^2\theta} (G_R(\omega_-) \mathcal{D}(\tilde{\sigma}_R^-)+G_R(-\omega_-) \mathcal{D}(\tilde{\sigma}_R^+))\nonumber\\&+\sin{^2\theta}(G_R(\omega_+)  \mathcal{D}(\tilde{\sigma}_L^-\tilde{\sigma}_R^z)+G_R(-\omega_+)  \mathcal{D}(\tilde{\sigma}_L^+\tilde{\sigma}_R^z)),
		\end{align}
		here, $\omega_\pm = \omega_{S}\pm\Omega$, and $\hat{\rho}(t)$ is the density matrix of the system of interest, and  $\tilde{\sigma}^{\pm}_i(\omega)$ are the jump operators in basis which diagonalize the system Hamiltonian. The explicit form of these jump operators is given in Appendix~\ref{section:Appendix A}.
		Furthermore, $G_{i}(\omega)$ denotes the spectral response function of the $i^{th}$ bath, and it is given by
		\begin{align}
		G_i(\omega)=
		\begin{cases}
		\gamma_{i}(\omega)(1+N_i(\omega) & \omega>0, \\
		\gamma_{i}(\omega)N_i(|\omega|) & \omega<0,
		\end{cases}
		\end{align}
		here, $N_i(\omega)$ is the Bose-Einstein distribution function (we take the Boltzmann constant as $k_B=1$),
		\begin{align}\label{eq:bose-einstein}
		N_i(\omega)=\frac{1}{e^{\omega/T_i}-1},
		\end{align}
		and coefficients $\gamma_{i}(\omega)$ are described by
		\begin{equation}
		\gamma_{i}(\omega) = 2\pi\hbar\frac{f_{i}(\omega)g_{i}(\omega)^2}{\omega},
		\end{equation}
		here $f_{i}(\omega)$ and $g_{i}(\omega)$ are density of modes of the baths and their interaction strengths to corresponding qubit, respectively. In the following, we consider thermal baths with flat density of modes, which makes $\gamma_{i}(\omega)$ independent of $\omega$ and can be denoted by $\gamma_{i}(\omega):=\kappa_{i}$. In the rest of the paper, we consider both baths have equal coupling strengths $\kappa_{L}=\kappa_{R}=\kappa$.   
		In Eq.~(\ref{Master}), the Lindblad dissipator for a jump operator $\hat A$ is defined by
		\begin{align}
		\mathcal{D}(\hat{A})=\hat{A}\hat{\rho}\hat{A}^{\dagger}-\frac{1}{2}(\hat{A}^{\dagger}\hat{A}\hat{\rho}+\hat{\rho}\hat{A}^{\dagger}\hat{A}).
		\end{align}
	We have ignored the dephasing term in Eq.~(\ref{Master}) because it does not influence the diagonal elements of the density matrix, consequently, it does not affect the steady-state heat currents~\cite{PhysRevE.85.061126}.
	We note that, in Eq.~(\ref{Master}) non-local jump operators are present, for example, $\tilde{\sigma}_L^z\tilde{\sigma}_R^-$, which means both baths have access to both qubits. Such a master equation is referred to global master equation which is consistent with the laws of thermodynamics~\cite{Levy_2014,PhysRevA.98.052123,Cattaneo_2019}. On contrary, a derivation based on neglecting the interaction term between the qubits leads to a local master equation which may not be consistent with the laws of thermodynamics~\cite{Levy_2014,PhysRevE.89.062109}.
	\section{Results} \label{section: Results}
	We find that heat rectification in our system can be characterized simply by using a single asymmetry parameter that can be introduced as
	\begin{align}\label{epsilon}
	\epsilon=\frac{|\omega_D|}{2g}   .
	\end{align}
	It measures the relative strength of detuning between the qubits compared to the DM interaction and controls the heat diode action. Significance of $\epsilon$ can be seen by expressing
	\begin{align}\label{cos-sine-define}
	\cos{^2\theta}&=\frac{1}{2\sqrt{1+\epsilon^2}(\sqrt{1+\epsilon^2}-\frac{|\omega_D|\epsilon}{\omega_D})}, \nonumber \\ \sin{^2\theta}&=\frac{\sqrt{1+\epsilon^2}-\frac{|\omega_D|\epsilon}{\omega_D}}{2\sqrt{1+\epsilon^2}}
	\end{align}
	and recognizing that $\epsilon$ controls the weight factors of different heat transfer channels described in Eq.~(\ref{Master}). Hence we deduce that $\epsilon$ can be envisioned as a valve to turn on or turn off
	various heat channels between the baths. For large values of $\epsilon$, depending on the sign of $\omega_D$ either $\cos{^2\theta}$ or $\sin{^2\theta}$ is very small and other tends to $1$, while for small values of $\epsilon$ both the function are almost equal and tend to $1/2$.
	\subsection{Heat flow analysis}\label{subsection::Current Results}
	Heat flux between a bath and the system is given by~\cite{Muhammad,e15062100}
	\begin{equation}\label{current_define}
	I_i=Tr[\mathcal{L}_{i}\hat{\rho}_S \tilde{H}_S],
	\end{equation} 	
	where $I_i$ represents the left $I_L$ or right bath $I_R$ current, with the sign convention of positive heat current if the heat flows from the bath to the system and vice versa. According to energy conservation, left and right steady-state heat currents must be the same but with opposite signs. Consequently, the evaluation of steady-state right bath current $I_R$ suffices for the qualitative and quantitative analysis of heat flow and thermal rectification. The right bath heat current $I_{R}$ evaluates to (see Appendix~\ref{App:Appendix:B} for details)
	\begin{widetext}
		\begin{align}\label{current_analytical}
		I_R(T_L,T_R)=&\ {\frac{\kappa}{4(1+\epsilon^2)}}\bigg[\frac{\omega_+(N_R(\omega_+)-N_L(\omega_+))}{D(T_L,T_R,\omega_+)}+\frac{\abs{\omega_-}(N_R(\abs{\omega_-})-N_L(\abs{\omega_-}))}{D(T_R,T_L,\abs{\omega_-})}\bigg],
		\end{align}
	\end{widetext}
	where,
	\begin{align}
	D(T_i,T_j,\omega)=\cos{^2\theta}(2 N_i(\omega)+1)+\sin{^2\theta}(2 N_j(\omega)+1),
	\end{align}
	and $I_R(T_L,T_R)$ indicates that the left bath temperature $T_L$ is greater than right bath temperature $T_R$ ( it is vice versa for $I_R(T_R,T_L)$). 
	
	
	\begin{figure}[!b] 
		\centering 
		\subfigure[]
		{\includegraphics[width=0.50\linewidth]{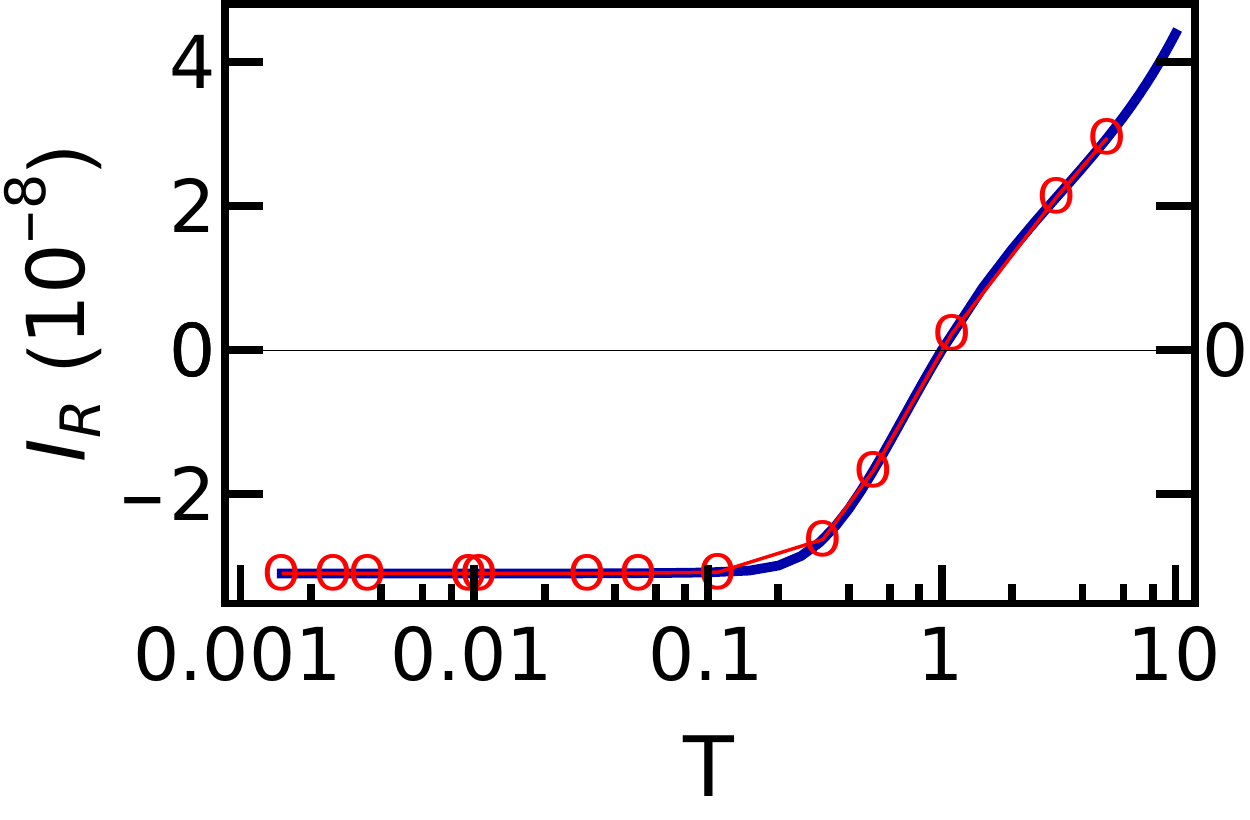}}\hfill
		\subfigure[]
		{\includegraphics[width=0.50\linewidth]{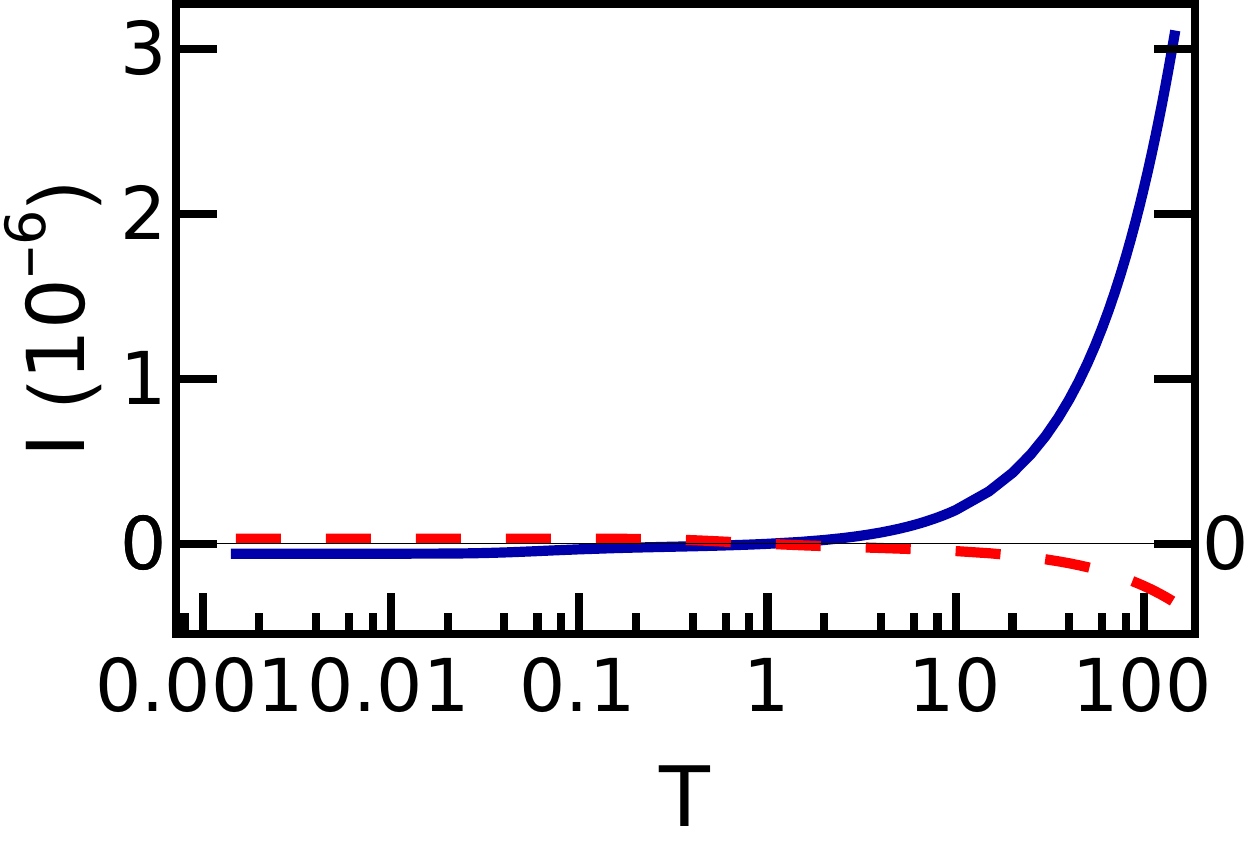}}
		\subfigure []
		{\includegraphics[width=0.50\linewidth]{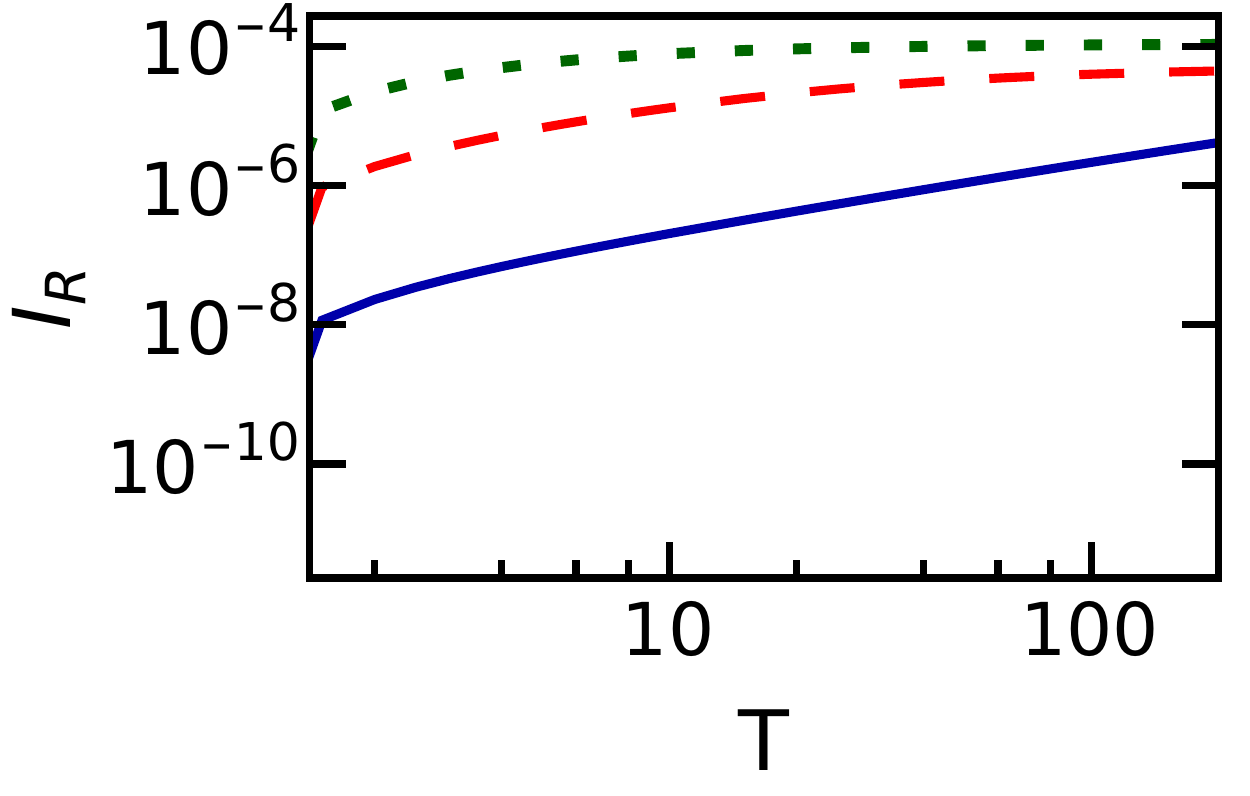}}\hfill
		\subfigure[]
		{\includegraphics[width=0.50\linewidth]{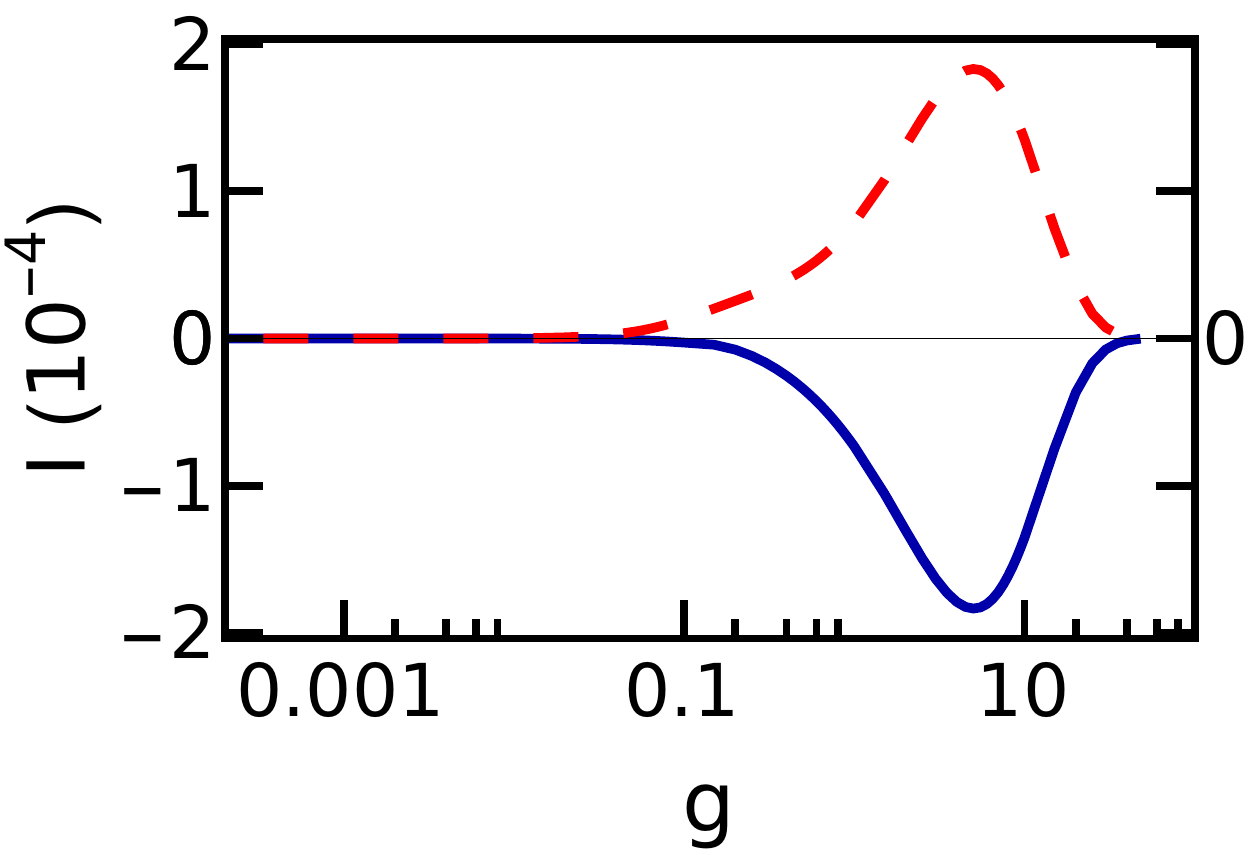}}
		\caption{(Color online) {Steady-state right heat bath current $I_{R}$ as a function of temperature $T$ and coupling strength $g$. (a) Comparison between the analytical (blue solid line) and numerical (red circles) results  of heat current $I_{R}$ evaluated via Eq.~(\ref{current_analytical}) and Eq.~(\ref{current_define}), respectively. We take $T_{\text{ref}}=1$ being reference temperature, and $T_R\equiv T,T_L=T_{\text{ref}}$ describes forward-biased configuration. (b) Forward-biased (blue solid line) and reverse-biased (red dashed line) heat current $I_{R}$, where reverse-biased (RB) configuration is described by $T_R=T_{\text{ref}},T_L\equiv T$. (c) Forward-biased (FB) heat current for  g=0.01 (blue solid line), g=0.1 (red dashed line), g=1.0 (green dotted line), and (d) $I_{R}$ as a function of DM interaction strength $g$ for RB current (blue solid line), and FB current (red dashed line). In both cases, we consider $T_{\text{High}}=10$, and $T_{\text{Low}}=1$. Rest of the parameters are given as $\omega_L=1$, $\omega_R=0.1$,  $\kappa=0.0001$, and $g=0.01$. All the parameters are scaled with the left qubit frequency $\omega_{L}=2\pi\times10$ GHz.}}
		\label{current_figure}
	\end{figure}
	

	In Eq.~(\ref{current_analytical}), the heat current $I_R$ depends on the square of DM interaction strength $g$, accordingly the direction of heat current is independent of the anti-symmetric nature of DM interaction. In Fig.~\ref{current_figure}, heat current $I_R$ is plotted as a function of temperature and coupling strength $g$. Since our model contains two heat baths, we set either one of the bath temperatures (left or right) as reference temperature and $T_{\text{ref}}=1$  unless otherwise specified. In Fig.~\ref{current_figure}(a), we verify the analytical result of heat current given in Eq.~(\ref{current_analytical}) by comparing it with the result obtained from the numerical solution of  Eq.~(\ref{Master}). From Fig.~\ref{current_figure}, we conclude that (i) The sign of heat current $I_R(T_L,T_R)$ is independent of the system parameters ($\omega_{S}$, $\Omega$, and $g$), it only changes with the interchange of bath temperatures. It is in accordance with the second law of thermodynamics. (ii) For $T_{R}>T_{L}$, heat current $I_R$ is positive, which indicates that heat current flows from right to left irrespective of the system parameters, and for $T_{L}>T_{R}$ it is vice versa. (iii) For weak coupling $g$, heat current $I_R(T_L,T_R)<I_R(T_R,T_L)$, which indicates that heat flow is suppressed from left to right, and it can be seen in Fig~\ref{current_figure}(b). (iv) Higher temperature gradients are associated with larger asymmetric heat flow (Fig.~\ref{current_figure}(c)). (v) Heat current vanishes for $g=0;\infty$, accordingly there exists a critical value of $g$ for which heat current is maximum.
	
	To examine the possible physical mechanism behind these observations, we refer to equation~\eqref{Master}. There are two heat transfer channels associated with the decay processes at $\omega_\pm$. For $g=0$, qubits are uncoupled, and there is no heat flow, which can be verified by Eq.~(\ref{current_analytical}). As we increase the coupling strength $g$, the dressed energy gap $\omega_{p}$ (Fig.~\ref{fig:engylev}) increases so that the phonon transfer channel $\omega_+$ acts at higher energy. Consequently, the heat current initially increases with $g$. However, once the energy levels are too far apart for the bath phonons to couple them, the current starts to decrease and eventually becomes zero. Taking the high temperature limit of \eqref{current_analytical} for the right bath, $T_R>>>(T_L,\omega_+)$), we get, 
	\begin{align}\label{current_limit}
	I_R(T_L,T_R)\approx&{\frac{\kappa}{4(1+\epsilon^2)}}T_R
	\nonumber\\&\bigg[\frac{1}{\cos{^2\theta}(2 N_L(\omega_+)+1)+\sin{^2\theta}(2\frac{T_R}{\omega_+})}+\nonumber\\&\frac{1}{\sin{^2\theta}(2 N_L(|\omega_-|)+1)+\cos{^2\theta}(2\frac{T_R}{|\omega_-|})}\bigg].
	\end{align} 
	This indicates that $I_R(T_L,T_R)$ linearly increases with $T_R$ and eventually saturates if  $\cos{^2\theta}$, and $\sin{^2\theta}$  are small. For weakly coupled qubits, either $\cos{^2\theta}$ or $\sin{^2\theta}$ is small. Consequently, heat flow saturates for larger temperature gradients. On the contrary, heat flow saturates at lower temperature gradients for larger coupling strength $g$, because both $\cos{^2\theta}$, and $\sin{^2\theta}$ have larger values in the strong coupling regime.  If the baths are sufficiently hot, the maximum saturation current we can derive in our system is
		\begin{align}\label{current_max}
		I_R^{\text{max}}(T_L,T_R)\propto\frac{\kappa}{4(1+\epsilon^2)}
		\end{align}
		We can see that the saturation current is larger for the case with {lower $\epsilon$}. We will see that this is in contrast with how the rectification behaves in the next section.
	
	\begin{figure}[!t]
		\begin{center}
			\leavevmode
			\includegraphics[width=0.47\textwidth,angle=0]{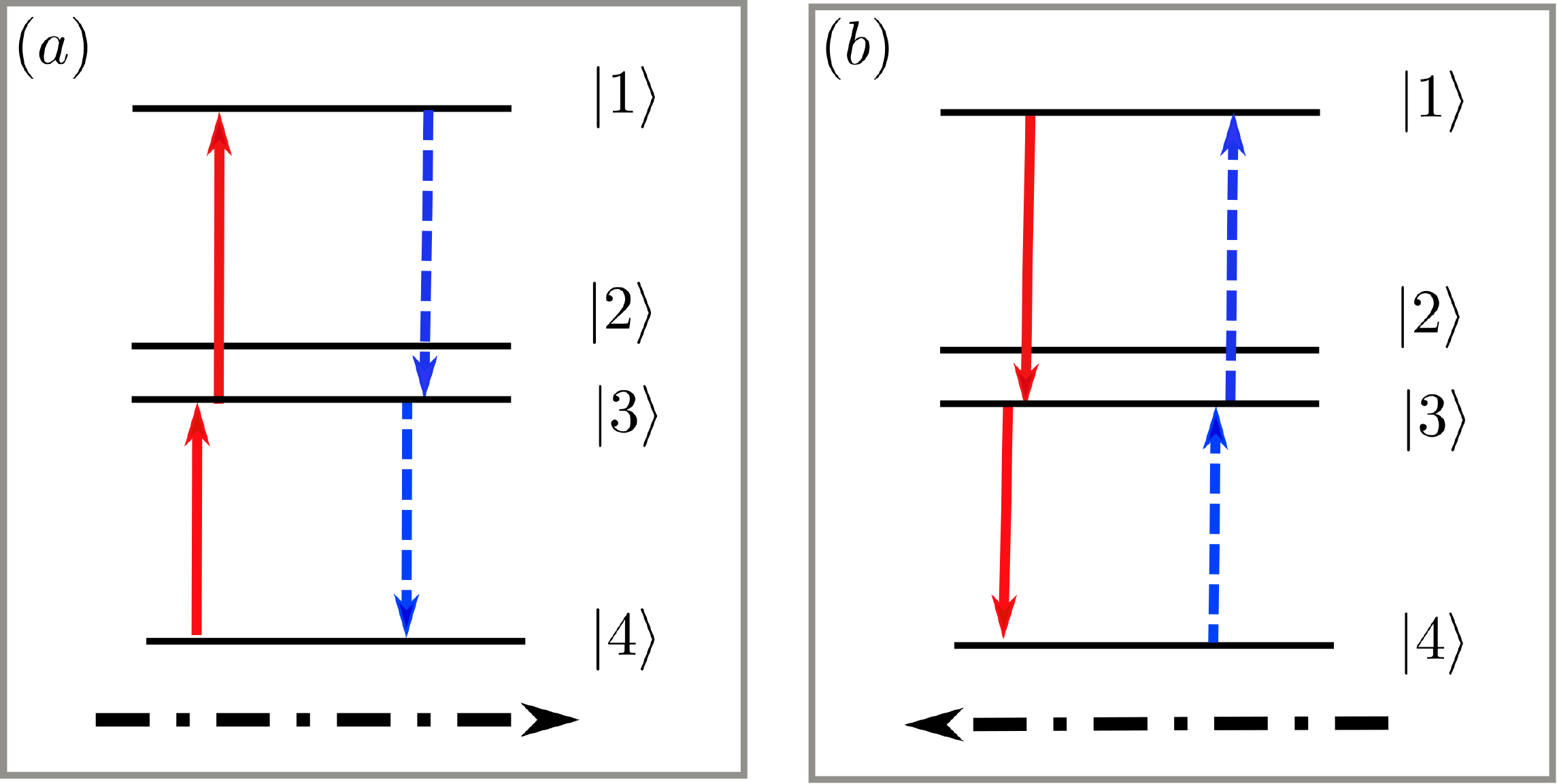}
			\caption{\label{fig:engylev2}{(Color online)  Examples of the processed that transfer heat between the baths for the weakly interacting resonant qubits $\omega_{L}=\omega_{R}=\omega\gg g$. The separation between energy levels $\ket{2}$, and $\ket{3}$ becomes $4g$, and $\omega_{\pm}$ transitions reduce to $\omega\pm 2g$. Solid, and dashed arrows indicate the transitions induced by the left and right baths, respectively, and the thickness of the arrows reflects the magnitudes of the decay rates between the states. In addition, dot-dashed arrows point the direction of the heat flow. For resonant qubits, all transition rates become symmetrical under the change in temperature bias due to which rectification becomes zero.}}
		\end{center}
	\end{figure}
	
	\subsection{Heat rectification}\label{subsection::Rectification Results}
	Out of the heat current results, it is straightforward to calculate the rectification factor, which is defined as
	\begin{equation}\label{eq:rectification}
	\mathcal{R}=\frac{ I_R(T_R,T_L)+I_R(T_L,T_R)}{\text{Max}[| I_R(T_L,T_R)|,| I_R(T_R,T_L) |]}.
	\end{equation}
	We note that here rectification factor $\mathcal{R}$ is based on the heat current $I_R$. However, identical results can be obtained by replacing $I_R$ with the left bath current $I_{L}$. The rectification factor $\mathcal{R}$ can take any value between -1 and 1, where $\mathcal{R}=1,-1$ describes perfect rectification, and $\mathcal{R}=0$ shows no asymmetry in the heat flow. In addition,  $\mathcal{R}>0$ means heat flow is suppressed from left to right, and $\mathcal{R}<0$ identifies the opposite case. To explain the physical mechanism behind rectification, we write the rate equation for population dynamics from the master equation~\eqref{Master} (for $\omega_s>\Omega$),
	{\fontsize{8pt}{12pt}
		\begin{align}\label{rate equation}
		\frac{d }{d t} \begin{bmatrix}
		\rho_{11}\\\rho_{22}
		\\\rho_{33}\\\rho_{44}
		\end{bmatrix}=\begin{pmatrix}
		-(r_3+r_4)& r_1& r_2&0\\r_3& -(r_1+r_4)& 0&r_2\\r_4& 0& -(r_2+r_3)&r_1\\0& r_4& r_3&-(r_1+r_2)
		\end{pmatrix}\begin{bmatrix}
		\rho_{11}\\\rho_{22}
		\\\rho_{33}\\\rho_{44}
		\end{bmatrix} 
		\end{align}}
	where $r_1, r_3$ are the transitions rates in $\omega_-$ channel and $r_2, r_4$ are the transitions rates in $\omega_+$ channel given as		
	\begin{align}\label{eq:decayrate}
	r_1&=\kappa\big[\sin{^2\theta} N_L(|\omega_-|)+\cos{^2\theta}N_R(|\omega_-|)\big],  \nonumber \\
	r_2&=\kappa\big[\cos{^2\theta} N_L(\omega_+)+\sin{^2\theta}N_R(\omega_+)\big],  \nonumber \\
	r_3&=\kappa\big[\sin{^2\theta} e^{\frac{|\omega_-|}{T_L}} N_L(|\omega_-|)+\cos{^2\theta} e^{\frac{|\omega_-|}{T_R}}N_R(|\omega_-|)\big],\nonumber\\
	r_4&=\kappa\big[\cos{^2\theta} e^{\frac{\omega_+}{T_L}} N_L(\omega_+)+\sin{^2\theta} e^{\frac{\omega_+}{T_R}}N_R(\omega_+)\big].
	\end{align}
	For  $\omega_s<\Omega$ we exchange $r_1$ and $r_3$ in the above equations. 
	Heat rectification in our model can be explained by possible four-wave mixing cycles responsible for heat flow between the left and right baths. In these cycles, the decay rates between two same dressed states of the qubits become significantly different when the thermal bias is reversed. Consequently, this causes an asymmetry in the heat flow. To elaborate more on this, let us look carefully at the rates given in equation \eqref{eq:decayrate}. The rates depend not only on the temperatures but also on $\cos{^2\theta}$, and $\sin{^2\theta}$, whose magnitudes [see Eq.~(\ref{cos-sine-define})] differ significantly for large $\epsilon$.  For appropriate system parameters, we can exploit this large dissimilarity to make some of the coupling strengths between the dressed states weaker than the others. 
These weak transitions can only be induced if coupled to a sufficiently hot bath. Accordingly, heat flow is suppressed in case of coupling the weak transitions with the weak field (cold bath)~\cite{Muhammad}, due to which heat flow has preferential direction in our model. As an example, for weakly coupled off-resonant qubits with $\omega_{D}\gg g> 0$, the transition rates associated with the left (right) bath $\omega_{-} (\omega_{+})$ decay channel becomes weaker because of the relative magnitude of $\cos{^2\theta}\gg\sin{^2\theta}$ [see Eqs.~(\ref{Master}) and (\ref{eq:decayrate})]. Consequently, for positively detuned qubits, $\omega_{-}$ channel is responsible for left to right heat flow suppression and it is vice versa for $\omega_{+}$. Hence, these two channels compete and have opposite signs in the rectification, which is given by
	\begin{align}\label{eq:rect}
	\mathcal{R}\propto I_R(T_R,T_L) + I_R(T_L,T_R) ,
	\end{align}
	\begin{align}\label{Rectification_analytical}
	\mathcal{R}\propto &{\frac{\frac{|\omega_D|}{\omega_D} \epsilon}{\sqrt{1+\epsilon^2}}}\bigg[\frac{\omega_+(N_R(\omega_+)-N_L(\omega_+))^2}{D(T_L,T_R,\omega_+)D(T_R,T_L,\omega_+)}\nonumber\\&-\frac{|\omega_-|(N_R(|\omega_-|)-N_L(|\omega_-|))^2}{D(T_R,T_L,|\omega_-|)D(T_L,T_R,|\omega_-|)}\bigg].
	\end{align}
Here, the first and second terms are associated with $\omega_{+}$ and $\omega_{-}$ channels, respectively. For $\omega_{D}>0$, if the right bath is cold, it may not be able to induce weak $\omega_{+}$ high energy transition. Accordingly, heat flow is suppressed from left to right, and rectification becomes positive due to the larger contribution of the first positive term compared to the second negative term in Eq.~(\ref{eq:rect}). This can also be verified from Eq.~(\ref{Rectification_analytical}), which shows that rectification is positive for $\omega_{D}>0$.
Similarly, negative detuning $\omega_{D}<0$ results in negative rectification, i.e., heat flow is suppressed from right to left.

	
	\begin{figure}[!t] 
		\centering 
		\subfigure[]
		{\includegraphics[width=0.50\linewidth]{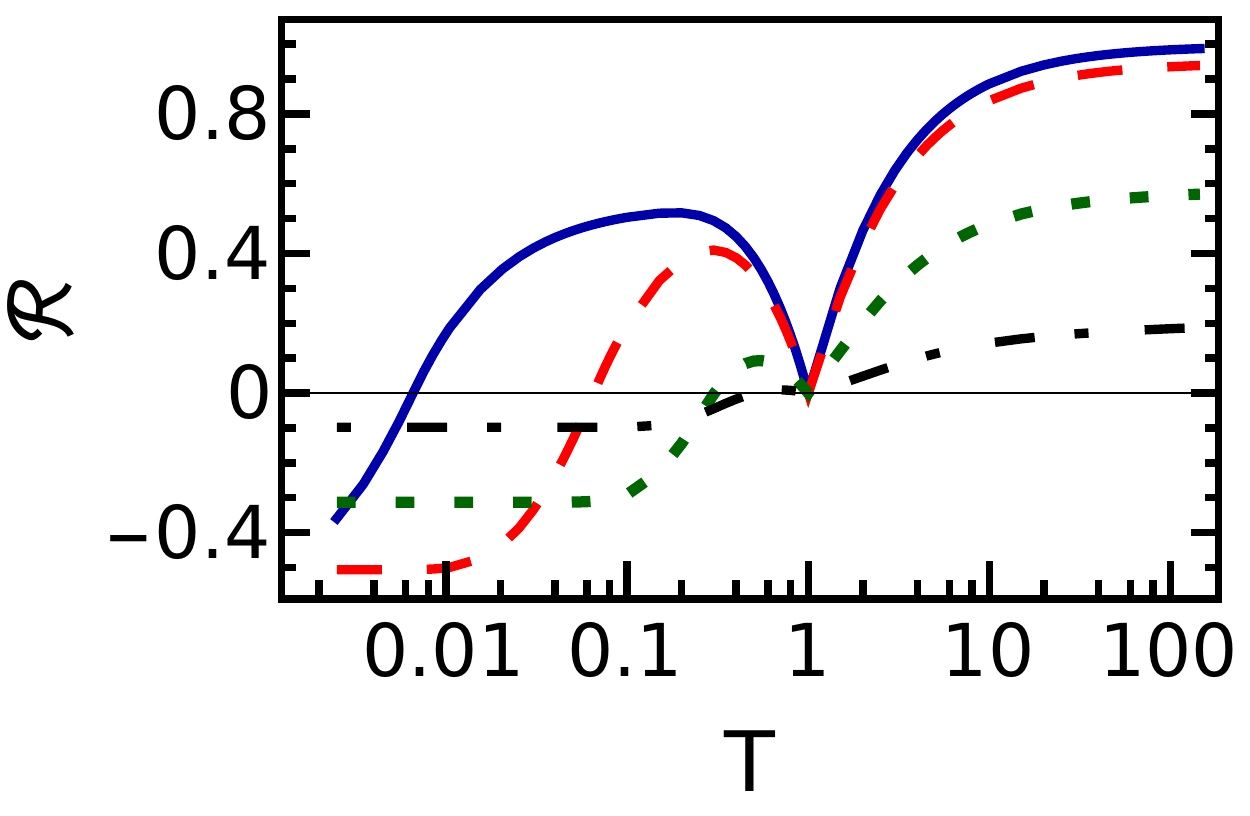}}\hfill
		\subfigure[]
		{\includegraphics[width=0.50\linewidth]{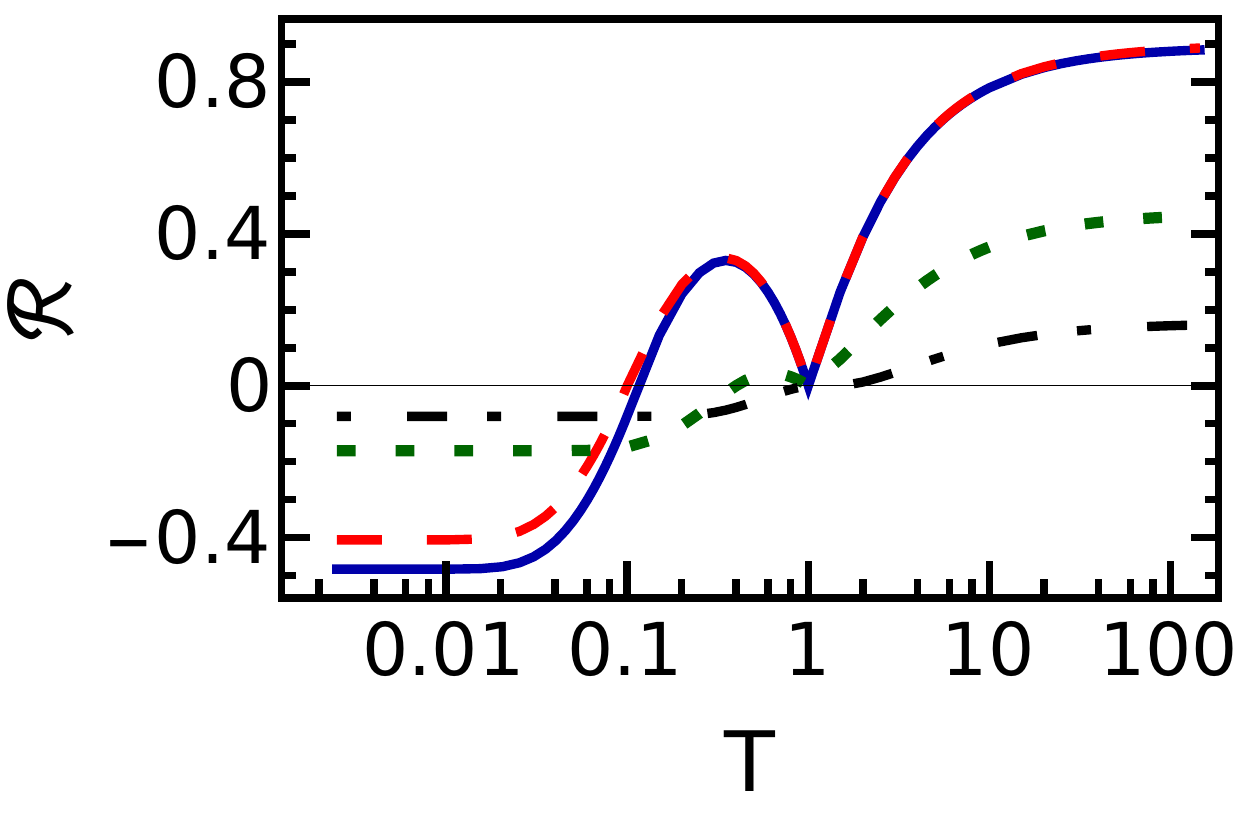}}
		\subfigure []
		{\includegraphics[width=0.50\linewidth]{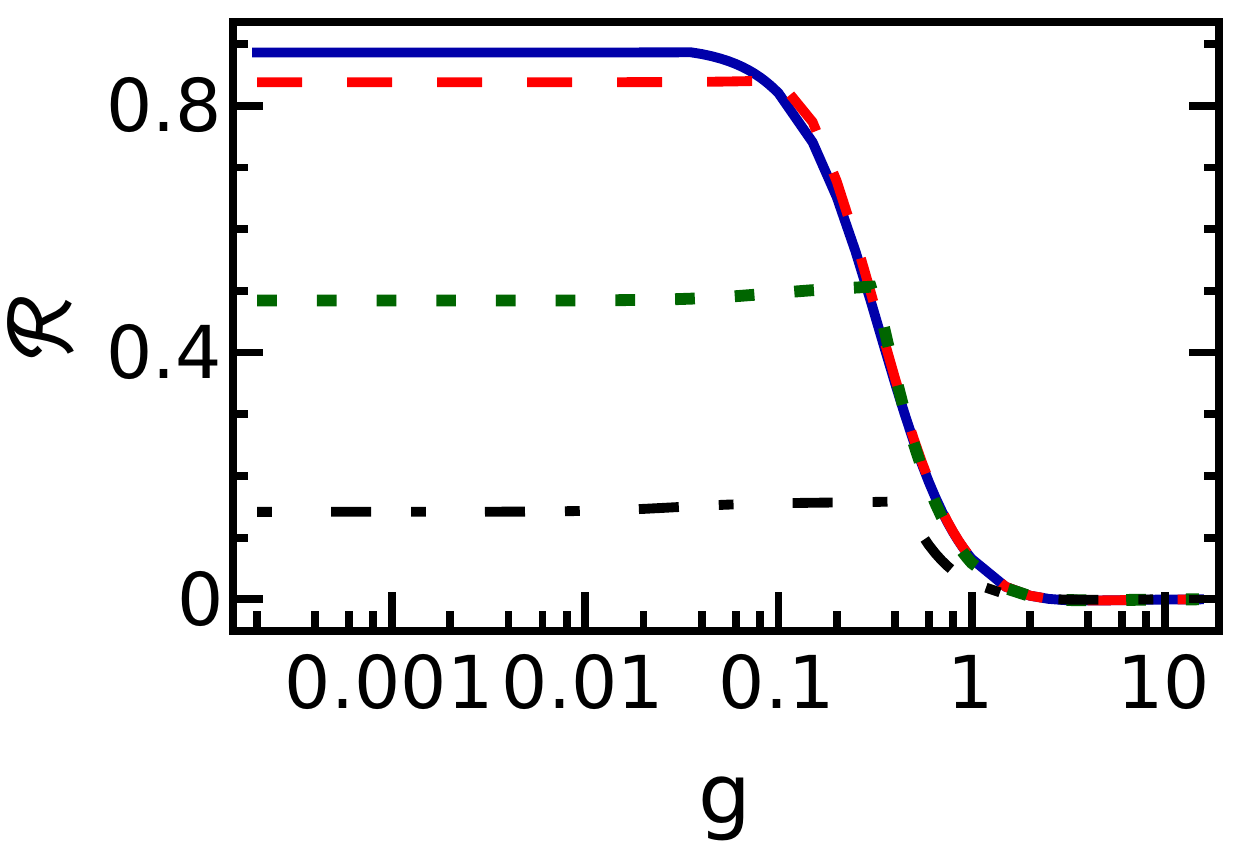}}\hfill
		\subfigure[]
		{\includegraphics[width=0.50\linewidth]{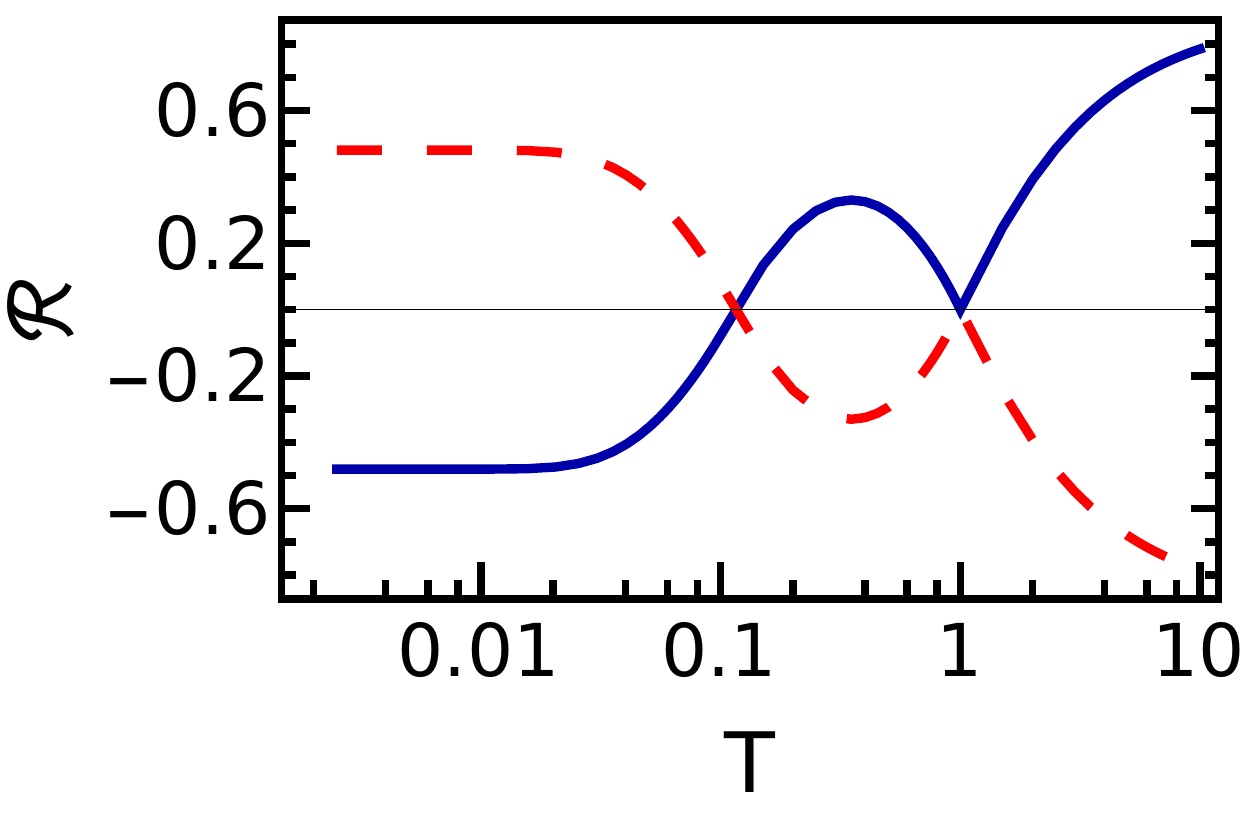}}
		\caption{(Color online) {Variation of rectification ($\mathcal{R}$) with temperature $T$ (top row)} for \textbf{(a)} $\omega_R$=0.005 (blue solid line), $\omega_R$=0.05 (red dashed line), $\omega_R$=0.4  (green dotted line), $\omega_R$=0.8  (black dot-dashed line), for \textbf{(b)} g=0.005 (blue solid line), g=0.05 (red dashed line), g=0.4  (green dotted line), g=0.8  (black dot-dashed line). Variation of $\mathcal{R}$ with $g$ for \textbf{(c)} $\omega_R$=0.005 (blue solid line), $\omega_R$=0.05 (red dashed line), $\omega_R$=0.4  (green dotted line), $\omega_R$=0.8  (black dot-dashed line) with $T_R=10$ and $T_L=1$.   \textbf{(d)} Shows the direction of rectification can be controlled by detuning $\omega_{D}$. Here, $\omega_D>0$ (blue solid line), $\omega_D<0$ (red dashed line). For all the cases, the values of parameters if not otherwise specified are $\omega_L= T_{\text{ref}}=1$, $\omega_R=0.1$, $g=0.01$, and $\kappa=0.0001$.}
		\label{Rectification_figure}
	\end{figure}
	
	For weakly interacting resonant qubits, i.e., $\omega_{L}=\omega_{R}=\omega\gg g$, an example of a process that transfers heat between the baths is shown in Fig.~\ref{fig:engylev2}. For resonant qubits, under the reversal of temperature gradient, the decay rates are given in Eq.~(\ref{eq:decayrate}) become invariant, due to which asymmetry in the heat flow vanishes. To emphasize this point, we explain the zero rectification for resonant qubits using Eq.~(\ref{eq:rectification}), heat flow becomes symmetric if
	\begin{align}
	I_R(T_L,T_R)=-I_R(T_R,T_L),
	\end{align}
	by simple manipulation this translates to
	\begin{align}
	D(T_L,T_R,\omega)=D(T_R,T_L,\omega),
	\end{align}
	which happens when
	\begin{align}\label{eq:AsymCond}
	\cos{^2\theta}-\sin{^2\theta}=0, \nonumber \\
	{\epsilon=0}.
	\end{align} 	
Here, we note that from  Eq.~\eqref{epsilon} this implies that the rectification is zero for resonant qubits. This is an expected result because we have already noted that the asymmetry of the cross product is not enough, and there is no other asymmetry in our model apart from the off-resonant qubits. Hence, there should be a direct relationship between the rectification and that off-resonance. 
	
Fig.~\ref{Rectification_figure} shows variations in the rectification $\mathcal{R}$ as a function of temperature $T$, and coupling strength $g$ for different detunings $\omega_{D}$. Higher rectification factors can be achieved for larger magnitudes of temperature gradients and detunings as shown in Fig.~\ref{Rectification_figure}(a). This is because according to Eq.~(\ref{Master}), the transition rates between the dressed states become more asymmetric for large detunings and temperature gradients. According to Eq.~(\ref{epsilon}), asymmetry in the heat flow decreases with the increase in the coupling strength $g$, and this is graphically represented in Figs.~\ref{Rectification_figure}(b) and \ref{Rectification_figure}(c). In our model, the direction of rectification can be controlled by the sign of detuning [see Eq.~(\ref{Rectification_analytical})], which is confirmed in Fig.~\ref{Rectification_figure}(d).	
	\begin{figure}[!t] 
		\centering 
		\subfigure[]
		{\includegraphics[width=0.50\linewidth]{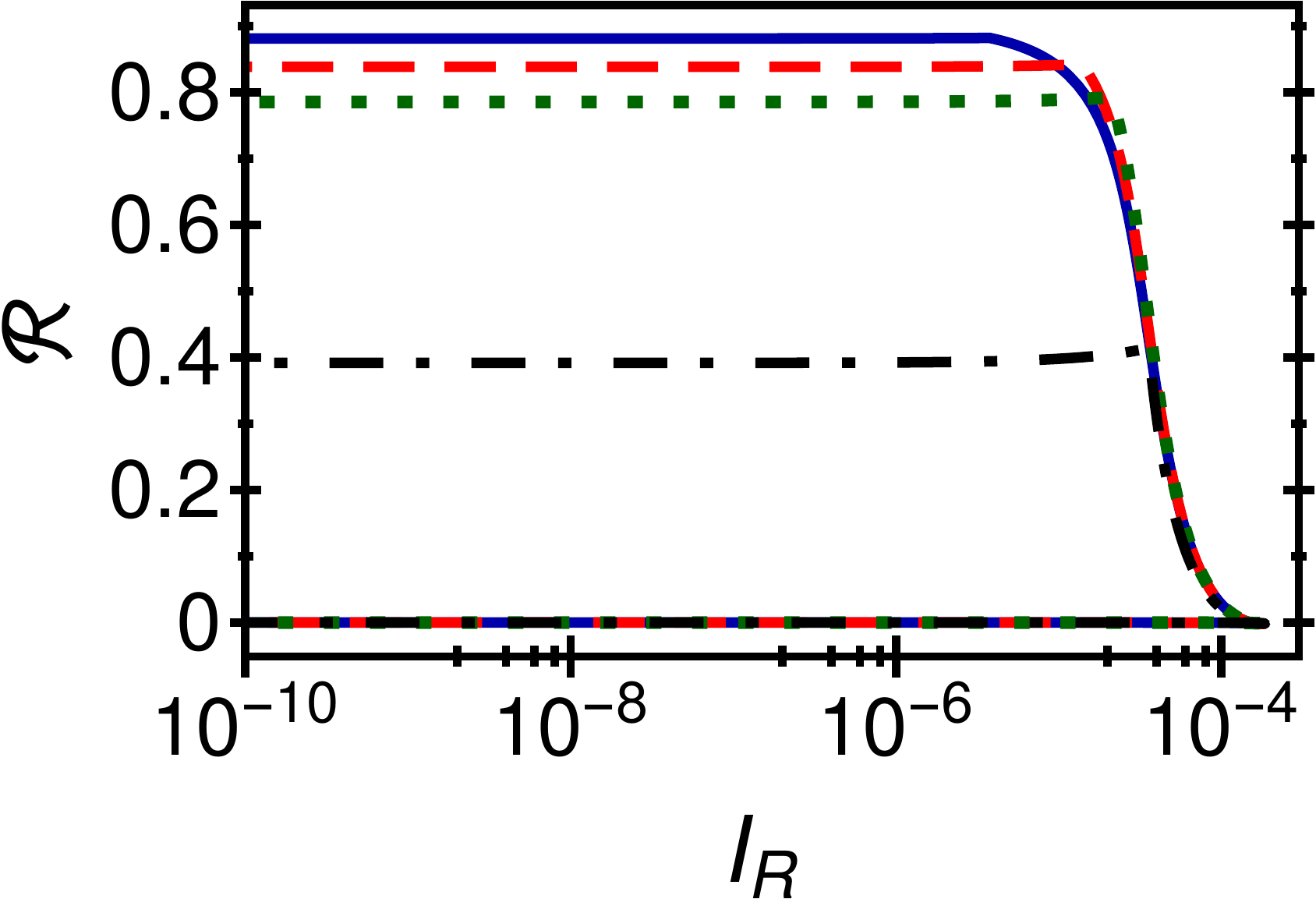}}\hfill
		\subfigure[]
		{\includegraphics[width=0.50\linewidth]{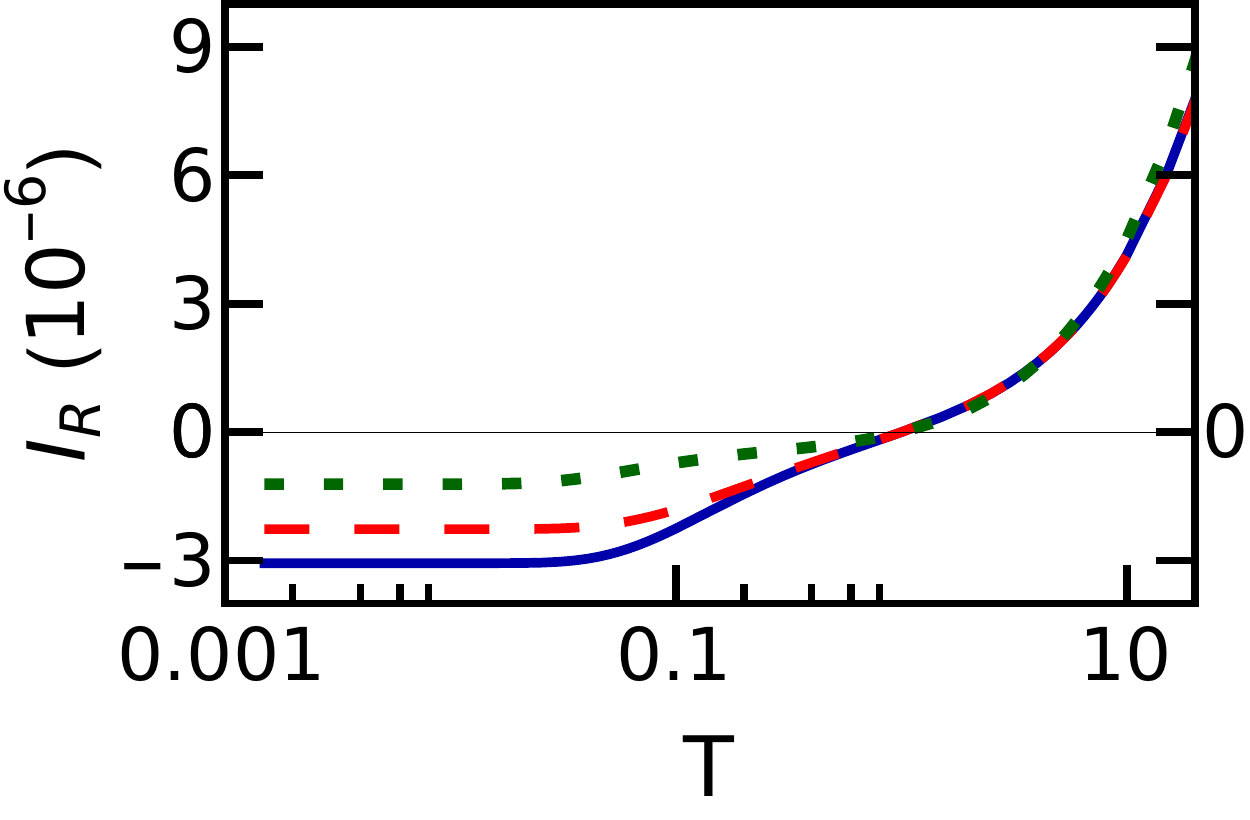}}
		\subfigure[]
		{\includegraphics[width=0.50\linewidth]{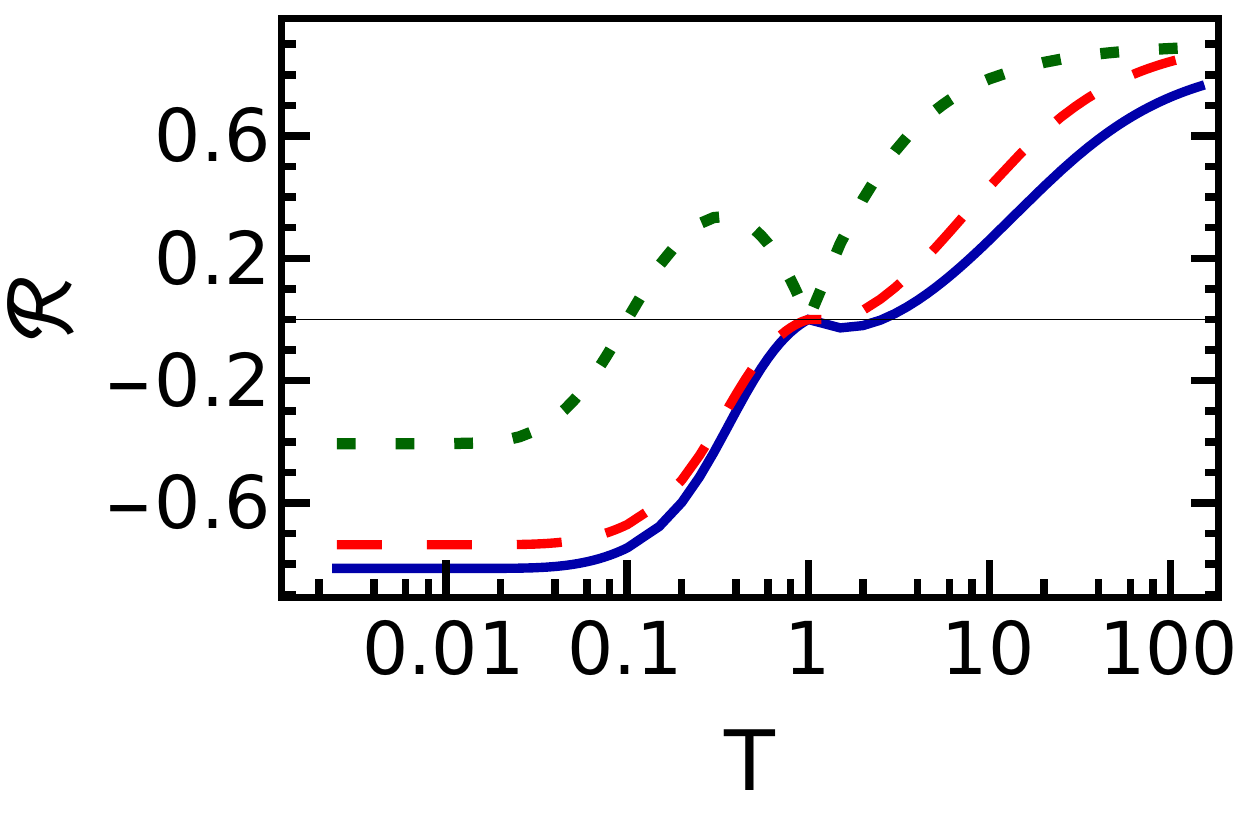}}\hfill
		\subfigure []
		{\includegraphics[width=0.50\linewidth]{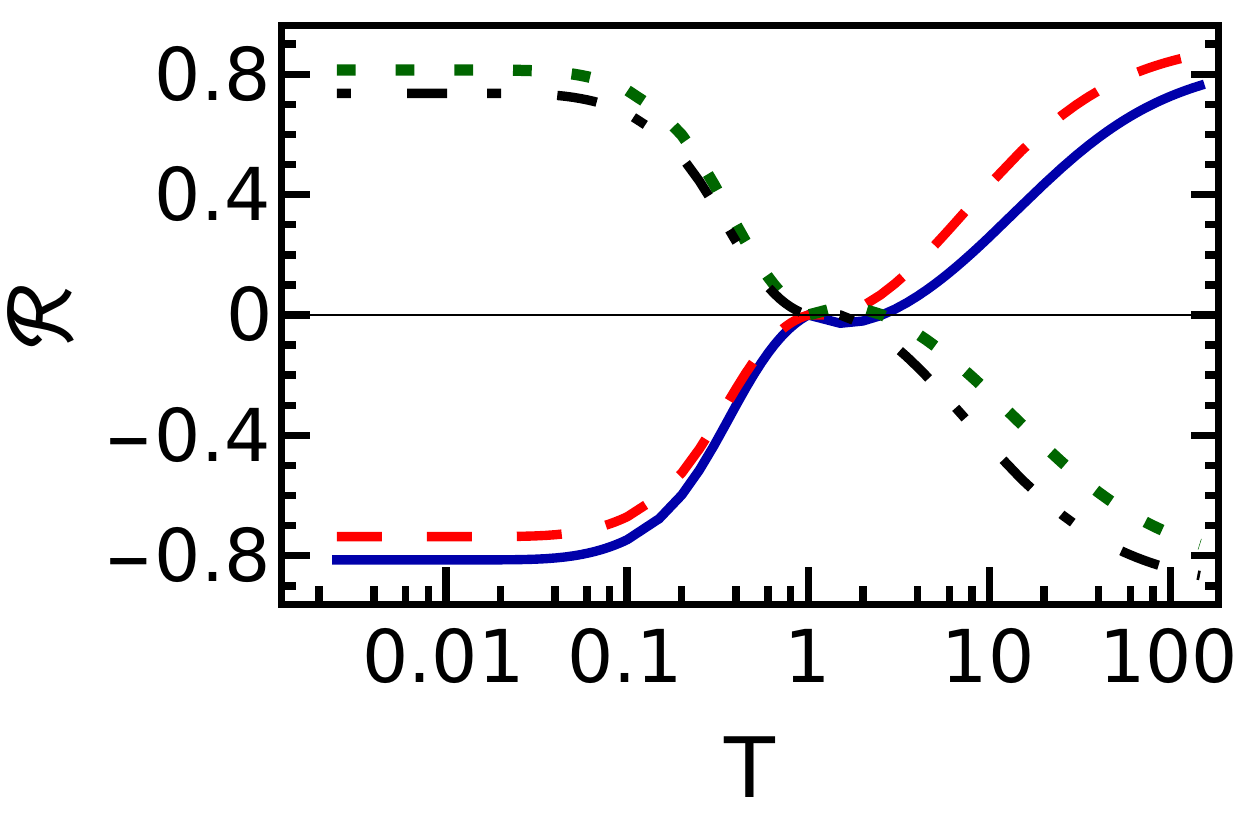}}
		\caption{\label{aniso}(Color online) {\textbf{(a)} Parametric curve between rectification ($\mathcal{R}$) 
				and $I_R$ for $\omega_R$=0.01 (blue solid line), $\omega_R$=0.05 (red dashed line), $\omega_R$=0.1 (green dotted line), $\omega_R$=0.5 (black dot-dashed line) with $T_R=10,$ $T_L=1$}. Variation of $I_R$ with $T$  for \textbf{(b)} DM along $x$ (blue solid line), along $y$ (red dashed line), along $z$ (green dotted line). Variation of 
			$\mathcal{R}$ with T    for \textbf{(c)} DM along $x$ (blue solid line), along $y$ (red dashed line), along $z$ 
			(green dotted line). \textbf{(d)} Changing the sign of $\mathcal{R}$ by exchange of qubit frequencies  for DM along $x$ (blue solid line, green dotted line) and for DM along $y$ (red dashed line, black dot-dashed line). For all the cases the values of parameters if not otherwise specified are $\omega_R=0.1$, $g=0.05$, $\kappa=0.0001$, 
			$\omega_L=1$, $T_{\text{ref}}=1.$ } 
		\label{anisotropy_figures}
	\end{figure}
	
Finally, from equation \eqref{Rectification_analytical} we see that the rectification is proportional to the constant
		\begin{align}\label{rectification_max}
		\mathcal{R}&\propto {\frac{\frac{|\omega_D|}{\omega_D} \epsilon}{\sqrt{1+\epsilon^2}}}
		\end{align}
		We have analytical results on the amount of detuning we need between the qubits in our model for getting significant rectification. We get high values of rectification when $\epsilon \to \infty$ or $\omega_D\gg2g$. Physically, this is so because the rectification arises due to the asymmetry in the energy levels of our system, determined by $\omega_D$; however, if $g$ is large compared to $\omega_D$, the asymmetry in energy levels is insufficient, and the rectification decreases. This is in line with the features we see in Fig.~\ref{Rectification_figure}.
		This constant also gives us an idea about the maximum rectification we can extract from our system, to  understand why that is so we recall that the two heat transfer channels work against each other in rectification  but for high values of rectification  the $\omega_-$ channel is of very low energy in comparison to the $\omega_+$  channel, hence the heat flow is completely dominated by the later. As a result, the rectification is also dominated by it. Looking again at the expression \eqref{Rectification_analytical}, we can see that if the $\omega_-$ channel is ignored the maximum achievable rectification is proportional to the constant given in \eqref{rectification_max}.
		Equations~\eqref{current_max} and~\eqref{rectification_max} reveal that there is a trade-off between the current and rectification as the saturation current is large for small $\epsilon$, whereas rectification diminishes, and vice versa. As pointed out earlier, this leads to a possibility of optimization, where we can get large heat currents without compromising the rectification, as can be seen in Fig. \ref{aniso}(a), where there is a region of stable rectification while current is increasing before sharply falling. Ideally, such a region should be targeted for the best performance of the thermal diode. The behavior of the curve further verifies that the saturation of rectification depends on the detuning between qubits. 
	
	\subsubsection{Effect of Anisotropy Field Direction}\label{subsection::Effect of anisotropy}
	
	From Fig. \ref{anisotropy_figures}(b,c,d), we can see that for low temperature regions ,the models containing the DM anisotropy field along $x$ and $y$ directions outperform the one with DM anisotropy field along $z$ direction in terms of both the current flow and rectification. This may be because there are more phonon transfer channels  available for these models as their Hamiltonian does not preserve total magnetisation. These channels are of relatively smaller energies allowing even the colder baths to induce sufficient transitions in them. However, again for higher temperature regions, the DM along $z$ model performs better.  We also see that the anisotropy field direction does not influence the fundamental  features in our diode as again the rectification changes sign on exchanging qubit frequencies and is zero for resonant qubits.
	
	\section{Quantumness of correlations and rectification}\label{section::coherences}
		
\begin{figure}[!htbp] 
	\centering 
	\subfigure[]
	{\includegraphics[width=0.8\linewidth]{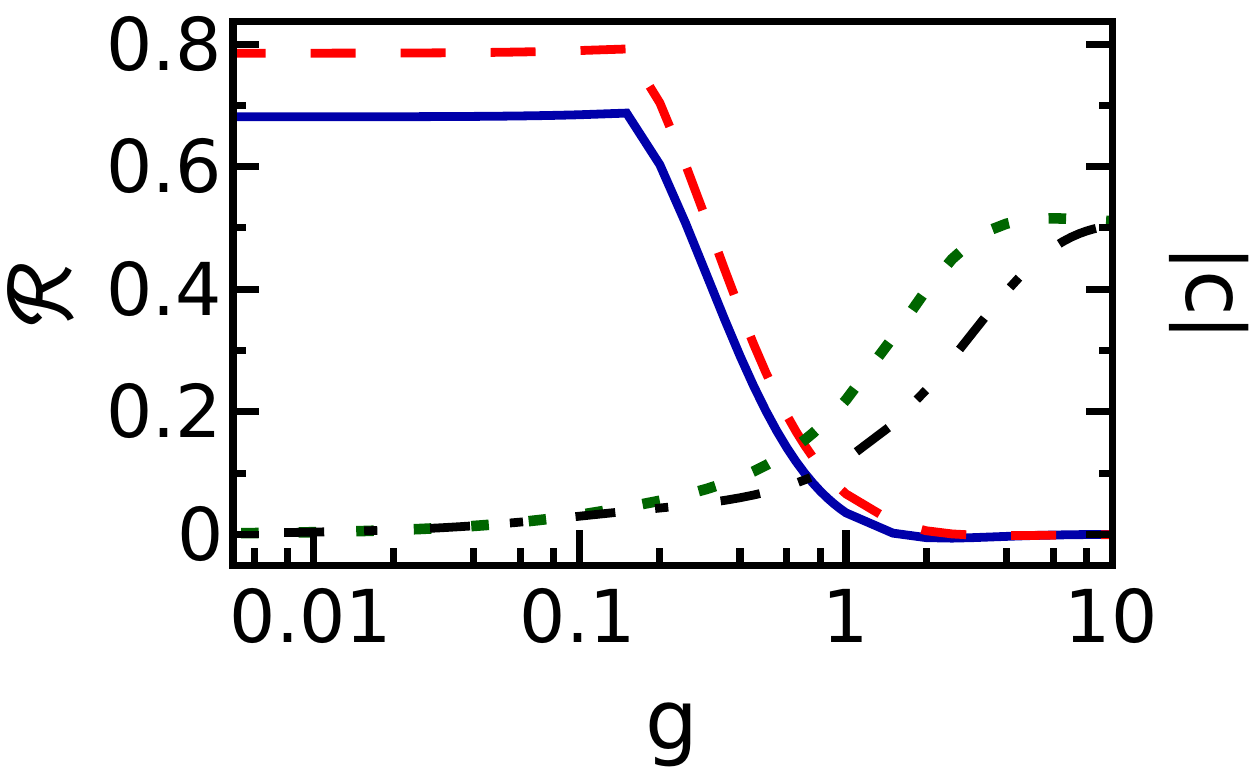}}
	\subfigure[]
	{\includegraphics[width=0.8\linewidth]{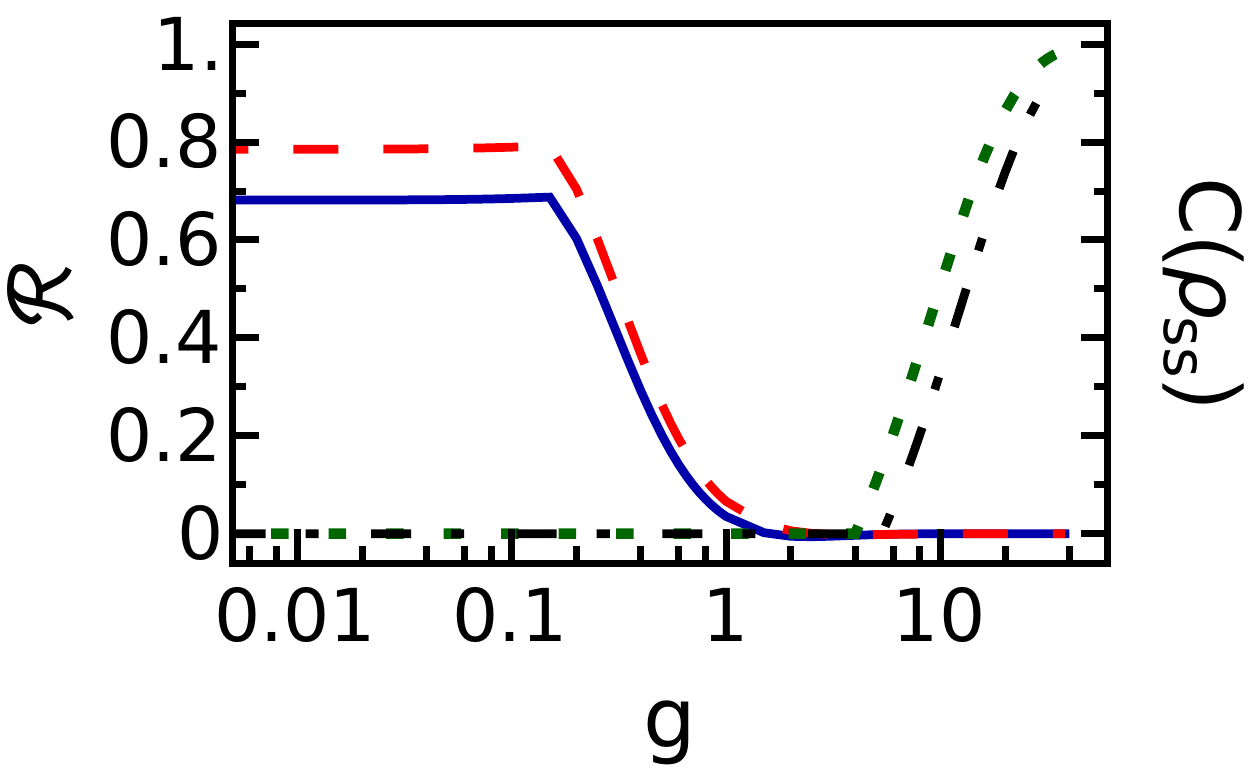}}
	\caption{\label{coherence_Fig} (Color online) {\textbf{(a)} Variation of rectification and coherence with $g$ for $T_R=5$ (blue solid line for $\mathcal{R}$) and (green dotted line for coherence) and $T_R=10$ (red dashed line for $\mathcal{R}$) and (black dot-dashed line for coherence) } and {\textbf{(b)}} Variation of rectification and concurrence with $g$ for $T_R$=5 (blue solid line for $\mathcal{R}$) and $($green dotted line for $C(\rho_{\text{ss}}))$ and $T_R$=10 $($red dashed line for $\mathcal{R}$) and $($black dot-dashed line $C(\rho_{\text{ss}}))$. The values of the parameters are $\omega_R=0.1$, $g=0.05$, $\kappa=0.0001$, $\omega_L=1$, $T_{\text{ref}}=1.$ }
	\label{coherence_fig}
\end{figure}

			Finally, we investigate any possible interplay between stationary quantum correlations and heat rectification in our model. The dissipative dynamics under Eq.~(\ref{Master}) imposes only two of the off-diagonal elements of the steady-state matrix in the computational basis remain non-zero. In the two qubits computational basis $\{|++\rangle, |+-\rangle, |-+\rangle, |--\rangle\}$, the steady-state density matrix $\rho_\text{ss}$ is given by a two-qubit X state (for $\omega_S>\Omega$)
		{\fontsize{8.5pt}{6pt}\begin{align}\label{computational_basis_matrix}
            \rho_\text{ss} =			
			\begin{pmatrix}
			d_1& 0& 0&0\\0&d_2& c &0\\0& c^{*}&d_3 &0\\0& 0& 0&d_4
			\end{pmatrix}. 
			\end{align}}
The steady-state diagonal elements (populations) are given by
\begin{align}
d_{1} & = \frac{r_1r_2}{D^{*}} & d_{2} = \frac{r_2r_3+r_1r_4+(r_2r_3-r_1r_4)\cos{(2\theta)}}{2 D^*}\\ \nonumber
d_{3} & = \frac{r_1r_2}{D^{*}} & d_{4} = \frac{r_2r_3+r_1r_4+(r_1r_4-r_2r_3)\cos{(2\theta)}}{2 D^*}, 
\end{align}			
For  $\omega_s<\Omega$, exchange $r_1$ and $r_3$ in the above equations. The off-diagonal term (coherence) is given by
\begin{align}\label{eq:c}
c = -i\frac{(r_2r_3-r_1r_4)\cos{\theta}\sin{\theta}}{D^*},
\end{align}
and for $ 4g^2<\omega_L\omega_R$ its absolute value is given by 
\begin{align}\label{eq:CohAsym}
 |c| & = \frac{\cos^2{\theta}(N^{-}_R-N^{+}_L)+\sin^2{\theta}(N^{-}_L-N^{+}_R)}{2 \sqrt{1+\epsilon^2}D^*}, \nonumber \\ &\text{ and for } 4g^2>\omega_L\omega_R\nonumber\\
|c| & = \frac{\cos^2{\theta}(N^{-}_R + N^{+}_L)+\sin^2{\theta}(N^{-}_L + N^{+}_R)+1}{2 \sqrt{1+\epsilon^2}D^*},
\end{align}
for convenience, we have used the following notations 
\begin{align}
N^{\pm}_{R(L)} & = N_{R(L)}(\omega_{\pm}) \nonumber \\ D^* &=D(T_R,T_L,|\omega_-|)D(T_L,T_R,\omega_+).
\end{align}	

As expected, for the uncoupled qubits, the coherences vanish because of $\text{cos}\theta=0$ in Eq.~(\ref{eq:c}), and the coherences are purely imaginary for any set of system parameters. For very large $g$ in comparison to $\omega_S$, $\omega_D$ and temperatures, the coherences saturate to
\begin{align}
|c|_{s} = \frac{1}{2\sqrt{1+\epsilon^2}},
\end{align}
which reflects that coherences are inversely proportional to $\epsilon$. We note that the coherence is necessary for the steady-state $\rho_\text{ss}$ to be in an entangled state; however, only coherence is not sufficient for its emergence. The precise condition for the two qubits to be in an entangled state is given by the positivity-of-the-partial-transpose separability criterion~\cite{PhysRevLett.77.1413}
\begin{align}
|c|>\frac{1}{2}\frac{d_{1}+d_4}{d_1-d_4}.
\end{align}
This condition is satisfied in the limit $g\gg\{\omega_{L}, \omega_{R}\}$.
To quantify the entanglement, we use concurrence as a measure of entanglement between the two qubits, and it is given by~\cite{Wootters} 
			\begin{equation}\label{con}
			C({\rho}_\text{ss})=\text{max}[0, ~\lambda_1-\lambda_2-\lambda_3-\lambda_4].
			\end{equation}
			Here $\lambda$'s are the eignevalue in decreasing order of the matrix
			\begin{equation}
			\hat{P}=\sqrt{\sqrt{{\rho}_\text{ss}} \tilde{\rho}_\text{ss}\sqrt{{\rho}_\text{ss}}}
			\end{equation}
			and 
			\begin{equation}
			\tilde{\rho}=(\hat{\sigma}_y \otimes \hat{\sigma}_y) {\rho}_\text{ss}^*(\hat{\sigma}_y \otimes \hat{\sigma}_y),
			\end{equation}
here complex conjugate operation is denoted by $*$.
			
 Fig.~\ref{coherence_fig} shows that both the stationary coherence $|c|$ and concurrence $C(\rho_{\text{ss}})$ are monotonically increasing functions of the interqubit coupling $g$, and both saturate to their maximum values in the limit $g\gg 1$. This is in contrast with the qualitative behavior of rectification, as larger value of $g$, is associated
with lower rectification [see Fig.~\ref{Rectification_figure}(c)]. Accordingly, strong quantum correlations and coherences are detrimental to the performance of our quantum thermal rectifier.
	
It is interesting to note that similar to heat currents, coherences are also asymmetrical under the reversal of temperature bias [see Eq.~(\ref{eq:CohAsym})]. Recall that the asymmetry in heat flow vanishes for $\epsilon = 0$ [given in Eq.~(\ref{eq:AsymCond})], which is possible for: (i) resonant qubits $\omega_{L}=\omega_{R}$, and (ii) $g\gg \{\omega_{L}, \omega_{R}\}$, in this limit $\text{sin}^2\theta \approx \text{cos}^2\theta$. Remarkably, the asymmetry in the coherences under the reversal of temperature bias also vanishes for these same conditions. Accordingly, similar to heat current rectification, we define asymmetry in coherences

\begin{align}\label{coherence_asymmetry}
\mathcal{A} :=\frac{ |c(T_R,T_L)|-|c(T_L,T_R)|}{\text{Max}[| c(T_L,T_R)|,| c(T_R,T_L) |]}.
\end{align}	
	
\begin{figure}[!t] 
	\centering 
	{\includegraphics[width=0.8\linewidth]{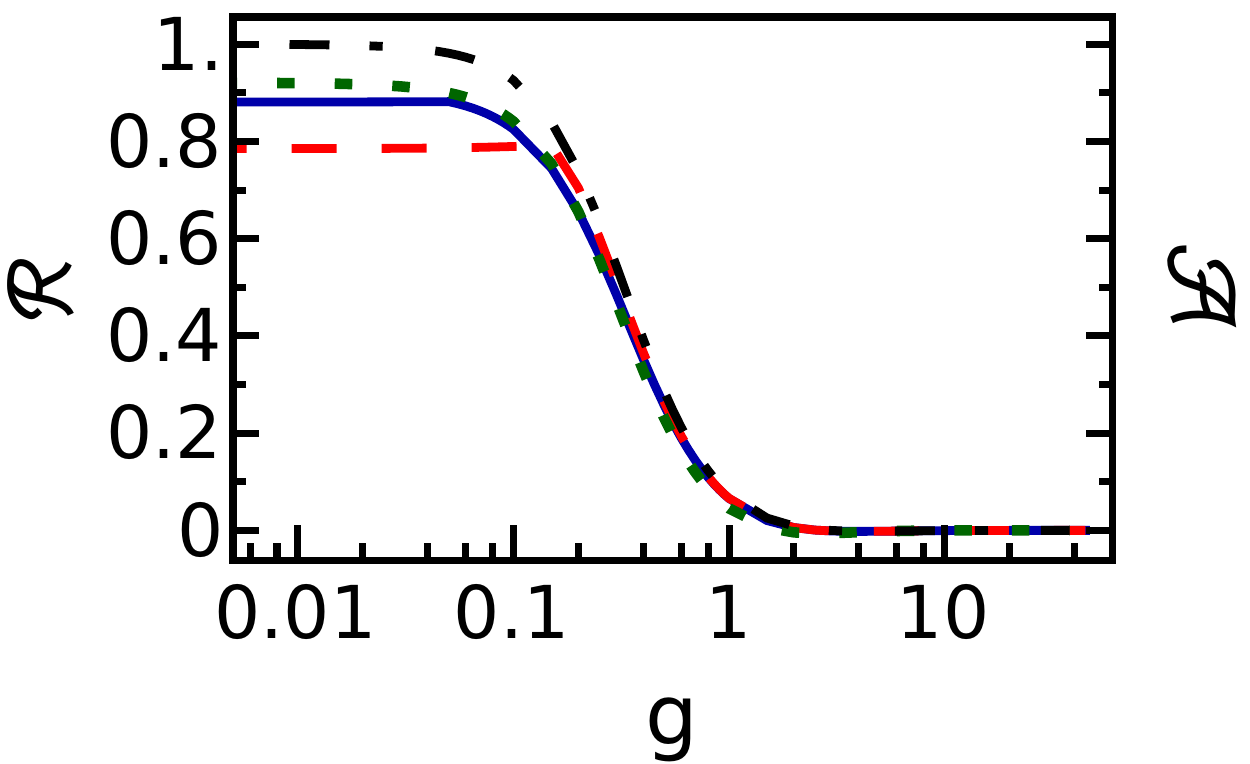}}
	\caption{\label{coherenceRect_Fig} (Color online) {\textbf{(a)}} Variation of rectification $\mathcal{R}$ and asymmetry in the coherences $\mathcal{A}$ as a function of $g$ for $T_\text{high}=5$ (blue solid line for $\mathcal{R}$) and (green dotted line for $\mathcal{A}$) and $T_\text{high}= 10$ (red dashed line for $\mathcal{R}$) and (black dot-dashed line $\mathcal{A}$).
		  Parameters: $\kappa=0.0001$,  $\omega_L=1$, and $\omega_{R}=0.01$.}
\end{figure}

To investigate the possible interplay between the asymmetry in the coherences and heat rectification, we plot both heat rectification $\mathcal{R}$ and asymmetry in coherence $\mathcal{A}$ as a function of the control parameter $g$ in Fig.~\ref{coherenceRect_Fig}.	
	In the limit of weak inter qubits coupling, $\mathcal{R}$ and $\mathcal{A}$ have similar qualitative behavior, and it becomes quantitatively identical as well for $g>\{\omega_{L}, \omega_{R}\}$. The increase in the temperature bias results in an increased asymmetry in the coherences. Accordingly, the quality of heat rectification also improves, which indicates that asymmetry in the coherences and heat rectification are associated. Asymmetry in the coherences is sufficient for the emergence of thermal rectification in our model.
	
	All the preceding analysis shows that the quantumness of correlations established between the qubits and heat rectification are related. We note that (i) the emergence of entanglement between the qubits kills the asymmetry in heat flow, and (ii) asymmetry in the coherence is required for the asymmetry in the heat flow and vice versa. 
	Along with the previously reported sources of asymmetry for thermal rectification~\cite{Muhammad}, we find that asymmetry in the coherences appears to be the fundamental resource for a thermal rectifier.

	\section{Conclusions}\label{sec:conclusion}
	
	We investigate the heat rectification ability of a two-qubit thermal diode in which the qubits interact via the DM interaction with the DM exchange field in the quantization axis. We find that thermal diode action is controlled by the relative strength of the detuning between the qubits compared to the DM field. We find that a single asymmetry parameter can be used to characterize the rectification in our system. We also see that there is a trade off between current and rectification in our system and larger current leads to a decrease in rectification but a possibility of optimization exists.
	We identify the high stability regions of the diode operation in terms of the bath temperatures and DM field amplitude. Furthermore, the direction of rectification can be controlled by the sign of the detuning of the qubits.  Similar features arise when we change the direction of the anisotropy field, though such DM thermal diodes operate more efficiently at lower temperatures. For higher temperatures, the DM exchange field along the quantization direction gives the optimum results.  
	
The heat rectification is found to be related to the stationary quantum correlations established between the qubits. The asymmetry in the heat flow vanishes with the emergence
of entanglement between the qubits. However, asymmetry in the coherences is found to be a fundamental resource for the performance of a quantum thermal rectifier. Correlation properties of the environment  may result in more efficient quantum thermal
diodes; however, it requires further investigation.
	
	\section{acknowledgement}
	
	RM gratefully acknowledges financial support from Science and Engineering Research Board (SERB), India, under the Core Research Grant (Project No. CRG/2020/000620). 
	We also thank Rafael S\'anchez of the Autonomous University of Madrid for his valuable feedback on our manuscript.
	\appendix
	
	\section{Master Equation}\label{section:Appendix A}
	
	The system Hamiltonian given in section \ref{eq:DMmodel} can be transformed into its diagonal basis using the transformation
	\begin{align}
	\hat{U}=\cos{^2(\frac{\theta}{2})} \hat{I}_L\otimes\hat{I}_R+\sin{^2(\frac{\theta}{2})}\hat{\sigma}_L^z\otimes \hat{\sigma}_R^z\nonumber\\+i \frac{\sin{\theta}}{2} (\hat{\sigma}_L^x\otimes\hat{\sigma}_R^x+\hat{\sigma}_L^y\otimes\hat{\sigma}_R^y).
	\end{align}
	The relation between the operators in new (dressed) basis with the operators in old(computational) basis is
	\begin{align}\label{transformation}
	\tilde{A}&=\hat{U}^{\dagger}\hat{A}\hat{U} \\
	\hat{A}&=\tilde{U}\tilde{A}\tilde{U}^{\dagger},
	\end{align}
	where $\tilde{A}$ is the operator in dressed basis and $\hat{A}$ is the operator in computational basis. It can be proved that under the transformation $\hat{U}$, the system Hamiltonian becomes a diagonal matrix with the following form in terms of dressed operators
	\begin{align}
	\tilde{H}_S=\frac{(\omega_S+\Omega)}{2} \tilde{\sigma}_L^z+\frac{(\omega_S-\Omega)}{2} \tilde{\sigma}_R^z
	\end{align} Now we want to write the system interaction operators in terms of these dressed operators. Using equation \eqref{transformation}, we have:
	\begin{align}
	\hat{\sigma}_i^x=\tilde{U}\tilde{\sigma}_i^x\tilde{U}^{\dagger},
	\end{align}
	with $i$ either left or right and
	\begin{align}
	\tilde{U}=\cos{^2(\frac{\theta}{2})} \tilde{I}_L\otimes\tilde{I}_R+\sin{^2(\frac{\theta}{2})}\tilde{\sigma}_L^z\otimes \tilde{\sigma}_R^z\nonumber\\+i \frac{\sin{\theta}}{2} (\tilde{\sigma}_L^x\otimes\tilde{\sigma}_R^x+\tilde{\sigma}_L^y\otimes\tilde{\sigma}_R^y).
	\end{align}
	Performing the above calculations, we get
	\begin{align}
	\hat{\sigma}_L^x=\cos{\theta}\tilde{\sigma}_L^x+\sin{\theta} \tilde{\sigma}_L^z\tilde{\sigma}_R^y\nonumber\\
	\hat{\sigma}_R^x=\cos{\theta}\tilde{\sigma}_R^x+\sin{\theta} \tilde{\sigma}_L^y\tilde{\sigma}_R^z.
	\end{align}
	The interaction picture version of these operators is ($e^{i\tilde{H}_S t}\tilde{A}e^{-i \tilde{H}_S t}$)
	
	\begin{align} \label{AppendixA_equation}
	\hat{\sigma}_L^x(t)&=e^{-i(\omega_S+\Omega)t}\cos{\theta}\tilde{\sigma}_L^-+ i e^{-i(\omega_S-\Omega)t}\sin{\theta}\tilde{\sigma}_L^z\tilde{\sigma}_R^-+H.C. \\
	\hat{\sigma}_R^x(t)&=e^{-i(\omega_S-\Omega)t}\cos{\theta}\tilde{\sigma}_R^-+ i e^{-i(\omega_S+\Omega)t}\sin{\theta}\tilde{\sigma}_L^-\tilde{\sigma}_R^z+H.C. 
	\end{align}
	
	Once we know the interaction picture operators, it is easy to arrive at the the Lindblad form of master equation given in \eqref{Master} \cite{breuer2002}.
	\section{Heat Current}\label{App:Appendix:B}
	
	Again from the master equation \eqref{Master}, we can see that
	\begin{align}
	\mathcal{L}_{R}\hat{\rho}(t)&=\cos{^2\theta} (G_R(\omega_-) \mathcal{D}(\tilde{\sigma}_R^-)+G_R(-\omega_-) \mathcal{D}(\tilde{\sigma}_R^+))\nonumber\\&+\sin{^2\theta}(G_R(\omega_+)  \mathcal{D}(\tilde{\sigma}_L^-\tilde{\sigma}_R^z)+G_R(-\omega_+) \mathcal{D}(\tilde{\sigma}_L^+\tilde{\sigma}_R^z)).
	\end{align}
	Using definition of current \eqref{current_define}, we get
	\begin{align}\label{current_expression}
	I_R&=\frac{\omega_+}{2}[\sin{^2\theta}(-G_R(\omega_+)  \langle I+\tilde{\sigma}_L^z \rangle +G_R(-\omega_+)  \langle I-\tilde{\sigma}_L^z \rangle)]\nonumber\\
	&+\frac{\omega_-}{2}[\cos{^2\theta}(-G_R(\omega_-)  \langle I+\tilde{\sigma}_R^z \rangle+G_R(-\omega_-)  \langle I-\tilde{\sigma}_R^z \rangle)].
	\end{align}
	We can find the dynamic equations of the average quantities required above
	\begin{align}
	\frac{d\tilde{\sigma}_L^z}{dt}=&\cos{^2\theta}(-G_L(\omega_+) \langle I+\tilde{\sigma}_L^z \rangle+G_L(-\omega_+) \langle I-\tilde{\sigma}_L^z \rangle)\nonumber\\
	&+\sin{^2 \theta}(-G_R(\omega_+) \langle I+\tilde{\sigma}_L^z \rangle
	+G_R(-\omega_+) \langle I-\tilde{\sigma}_L^z \rangle), \\
	\frac{d\tilde{\sigma}_R^z}{dt}=&\sin{^2\theta}(-G_L(\omega_-) \langle I+\tilde{\sigma}_R^z \rangle
	+G_L(-\omega_-)   \langle I-\tilde{\sigma}_R^z \rangle)\nonumber\\
	&+\cos{^2 \theta}(-G_R(\omega_-) \langle I+\tilde{\sigma}_R^z \rangle
	+G_R(-\omega_-) \langle I-\tilde{\sigma}_R^z \rangle).
	\end{align}
	For finding the steady state solution the L.H.S. of both the equations above goes to zero. Using the steady state value of the above averages and putting it in the expression of current \eqref{current_expression} gives us the analytical form of the current in \eqref{current_analytical}.
	
	\bibliography{diode}

\providecommand{\noopsort}[1]{}\providecommand{\singleletter}[1]{#1}%
\begin{thebibliography}{71}%
\makeatletter
\providecommand \@ifxundefined [1]{%
 \@ifx{#1\undefined}
}%
\providecommand \@ifnum [1]{%
 \ifnum #1\expandafter \@firstoftwo
 \else \expandafter \@secondoftwo
 \fi
}%
\providecommand \@ifx [1]{%
 \ifx #1\expandafter \@firstoftwo
 \else \expandafter \@secondoftwo
 \fi
}%
\providecommand \natexlab [1]{#1}%
\providecommand \enquote  [1]{``#1''}%
\providecommand \bibnamefont  [1]{#1}%
\providecommand \bibfnamefont [1]{#1}%
\providecommand \citenamefont [1]{#1}%
\providecommand \href@noop [0]{\@secondoftwo}%
\providecommand \href [0]{\begingroup \@sanitize@url \@href}%
\providecommand \@href[1]{\@@startlink{#1}\@@href}%
\providecommand \@@href[1]{\endgroup#1\@@endlink}%
\providecommand \@sanitize@url [0]{\catcode `\\12\catcode `\$12\catcode
  `\&12\catcode `\#12\catcode `\^12\catcode `\_12\catcode `\%12\relax}%
\providecommand \@@startlink[1]{}%
\providecommand \@@endlink[0]{}%
\providecommand \url  [0]{\begingroup\@sanitize@url \@url }%
\providecommand \@url [1]{\endgroup\@href {#1}{\urlprefix }}%
\providecommand \urlprefix  [0]{URL }%
\providecommand \Eprint [0]{\href }%
\providecommand \doibase [0]{http://dx.doi.org/}%
\providecommand \selectlanguage [0]{\@gobble}%
\providecommand \bibinfo  [0]{\@secondoftwo}%
\providecommand \bibfield  [0]{\@secondoftwo}%
\providecommand \translation [1]{[#1]}%
\providecommand \BibitemOpen [0]{}%
\providecommand \bibitemStop [0]{}%
\providecommand \bibitemNoStop [0]{.\EOS\space}%
\providecommand \EOS [0]{\spacefactor3000\relax}%
\providecommand \BibitemShut  [1]{\csname bibitem#1\endcsname}%
\let\auto@bib@innerbib\@empty
\bibitem [{\citenamefont {Terraneo}\ \emph {et~al.}(2002)\citenamefont
  {Terraneo}, \citenamefont {Peyrard},\ and\ \citenamefont
  {Casati}}]{PhysRevLett.88.094302}%
  \BibitemOpen
  \bibfield  {author} {\bibinfo {author} {\bibfnamefont {M.}~\bibnamefont
  {Terraneo}}, \bibinfo {author} {\bibfnamefont {M.}~\bibnamefont {Peyrard}}, \
  and\ \bibinfo {author} {\bibfnamefont {G.}~\bibnamefont {Casati}},\ }\href
  {\doibase 10.1103/PhysRevLett.88.094302} {\bibfield  {journal} {\bibinfo
  {journal} {Phys. Rev. Lett.}\ }\textbf {\bibinfo {volume} {88}},\ \bibinfo
  {pages} {094302} (\bibinfo {year} {2002})}\BibitemShut {NoStop}%
\bibitem [{\citenamefont {Lepri}\ \emph {et~al.}(2003)\citenamefont {Lepri},
  \citenamefont {Livi},\ and\ \citenamefont {Politi}}]{classical1}%
  \BibitemOpen
  \bibfield  {author} {\bibinfo {author} {\bibfnamefont {S.}~\bibnamefont
  {Lepri}}, \bibinfo {author} {\bibfnamefont {R.}~\bibnamefont {Livi}}, \ and\
  \bibinfo {author} {\bibfnamefont {A.}~\bibnamefont {Politi}},\ }\href
  {\doibase https://doi.org/10.1016/S0370-1573(02)00558-6} {\bibfield
  {journal} {\bibinfo  {journal} {Phys. Rep.}\ }\textbf {\bibinfo {volume}
  {377}},\ \bibinfo {pages} {1} (\bibinfo {year} {2003})}\BibitemShut {NoStop}%
\bibitem [{\citenamefont {Li}\ \emph {et~al.}(2004)\citenamefont {Li},
  \citenamefont {Wang},\ and\ \citenamefont {Casati}}]{PhysRevLett.93.184301}%
  \BibitemOpen
  \bibfield  {author} {\bibinfo {author} {\bibfnamefont {B.}~\bibnamefont
  {Li}}, \bibinfo {author} {\bibfnamefont {L.}~\bibnamefont {Wang}}, \ and\
  \bibinfo {author} {\bibfnamefont {G.}~\bibnamefont {Casati}},\ }\href
  {\doibase 10.1103/PhysRevLett.93.184301} {\bibfield  {journal} {\bibinfo
  {journal} {Phys. Rev. Lett.}\ }\textbf {\bibinfo {volume} {93}},\ \bibinfo
  {pages} {184301} (\bibinfo {year} {2004})}\BibitemShut {NoStop}%
\bibitem [{\citenamefont {Segal}\ and\ \citenamefont
  {Nitzan}(2005)}]{spin_boson_thermal_rectifier}%
  \BibitemOpen
  \bibfield  {author} {\bibinfo {author} {\bibfnamefont {D.}~\bibnamefont
  {Segal}}\ and\ \bibinfo {author} {\bibfnamefont {A.}~\bibnamefont {Nitzan}},\
  }\href {\doibase 10.1103/PhysRevLett.94.034301} {\bibfield  {journal}
  {\bibinfo  {journal} {Phys. Rev. Lett.}\ }\textbf {\bibinfo {volume} {94}},\
  \bibinfo {pages} {034301} (\bibinfo {year} {2005})}\BibitemShut {NoStop}%
\bibitem [{\citenamefont {Lan}\ and\ \citenamefont
  {Li}(2006)}]{PhysRevB.74.214305}%
  \BibitemOpen
  \bibfield  {author} {\bibinfo {author} {\bibfnamefont {J.}~\bibnamefont
  {Lan}}\ and\ \bibinfo {author} {\bibfnamefont {B.}~\bibnamefont {Li}},\
  }\href {\doibase 10.1103/PhysRevB.74.214305} {\bibfield  {journal} {\bibinfo
  {journal} {Phys. Rev. B}\ }\textbf {\bibinfo {volume} {74}},\ \bibinfo
  {pages} {214305} (\bibinfo {year} {2006})}\BibitemShut {NoStop}%
\bibitem [{\citenamefont {Scheibner}\ \emph {et~al.}(2008)\citenamefont
  {Scheibner}, \citenamefont {König}, \citenamefont {Reuter}, \citenamefont
  {Wieck}, \citenamefont {Gould}, \citenamefont {Buhmann},\ and\ \citenamefont
  {Molenkamp}}]{Scheibner_2008}%
  \BibitemOpen
  \bibfield  {author} {\bibinfo {author} {\bibfnamefont {R.}~\bibnamefont
  {Scheibner}}, \bibinfo {author} {\bibfnamefont {M.}~\bibnamefont {König}},
  \bibinfo {author} {\bibfnamefont {D.}~\bibnamefont {Reuter}}, \bibinfo
  {author} {\bibfnamefont {A.~D.}\ \bibnamefont {Wieck}}, \bibinfo {author}
  {\bibfnamefont {C.}~\bibnamefont {Gould}}, \bibinfo {author} {\bibfnamefont
  {H.}~\bibnamefont {Buhmann}}, \ and\ \bibinfo {author} {\bibfnamefont
  {L.~W.}\ \bibnamefont {Molenkamp}},\ }\href {\doibase
  10.1088/1367-2630/10/8/083016} {\bibfield  {journal} {\bibinfo  {journal}
  {New J. Phys.}\ }\textbf {\bibinfo {volume} {10}},\ \bibinfo {pages} {083016}
  (\bibinfo {year} {2008})}\BibitemShut {NoStop}%
\bibitem [{\citenamefont {Segal}(2008)}]{PhysRevLett.100.105901}%
  \BibitemOpen
  \bibfield  {author} {\bibinfo {author} {\bibfnamefont {D.}~\bibnamefont
  {Segal}},\ }\href {\doibase 10.1103/PhysRevLett.100.105901} {\bibfield
  {journal} {\bibinfo  {journal} {Phys. Rev. Lett.}\ }\textbf {\bibinfo
  {volume} {100}},\ \bibinfo {pages} {105901} (\bibinfo {year}
  {2008})}\BibitemShut {NoStop}%
\bibitem [{\citenamefont {Wu}\ \emph {et~al.}(2009)\citenamefont {Wu},
  \citenamefont {Yu},\ and\ \citenamefont {Segal}}]{PhysRevE.80.041103}%
  \BibitemOpen
  \bibfield  {author} {\bibinfo {author} {\bibfnamefont {L.-A.}\ \bibnamefont
  {Wu}}, \bibinfo {author} {\bibfnamefont {C.~X.}\ \bibnamefont {Yu}}, \ and\
  \bibinfo {author} {\bibfnamefont {D.}~\bibnamefont {Segal}},\ }\href
  {\doibase 10.1103/PhysRevE.80.041103} {\bibfield  {journal} {\bibinfo
  {journal} {Phys. Rev. E}\ }\textbf {\bibinfo {volume} {80}},\ \bibinfo
  {pages} {041103} (\bibinfo {year} {2009})}\BibitemShut {NoStop}%
\bibitem [{\citenamefont {Ruokola}\ \emph {et~al.}(2009)\citenamefont
  {Ruokola}, \citenamefont {Ojanen},\ and\ \citenamefont
  {Jauho}}]{PhysRevB.79.144306}%
  \BibitemOpen
  \bibfield  {author} {\bibinfo {author} {\bibfnamefont {T.}~\bibnamefont
  {Ruokola}}, \bibinfo {author} {\bibfnamefont {T.}~\bibnamefont {Ojanen}}, \
  and\ \bibinfo {author} {\bibfnamefont {A.-P.}\ \bibnamefont {Jauho}},\ }\href
  {\doibase 10.1103/PhysRevB.79.144306} {\bibfield  {journal} {\bibinfo
  {journal} {Phys. Rev. B}\ }\textbf {\bibinfo {volume} {79}},\ \bibinfo
  {pages} {144306} (\bibinfo {year} {2009})}\BibitemShut {NoStop}%
\bibitem [{\citenamefont {Wu}\ and\ \citenamefont
  {Segal}(2009)}]{PhysRevLett.102.095503}%
  \BibitemOpen
  \bibfield  {author} {\bibinfo {author} {\bibfnamefont {L.-A.}\ \bibnamefont
  {Wu}}\ and\ \bibinfo {author} {\bibfnamefont {D.}~\bibnamefont {Segal}},\
  }\href {\doibase 10.1103/PhysRevLett.102.095503} {\bibfield  {journal}
  {\bibinfo  {journal} {Phys. Rev. Lett.}\ }\textbf {\bibinfo {volume} {102}},\
  \bibinfo {pages} {095503} (\bibinfo {year} {2009})}\BibitemShut {NoStop}%
\bibitem [{\citenamefont {Zhang}\ \emph {et~al.}(2009)\citenamefont {Zhang},
  \citenamefont {Yan}, \citenamefont {Wu}, \citenamefont {Wang},\ and\
  \citenamefont {Li}}]{PhysRevB.80.172301}%
  \BibitemOpen
  \bibfield  {author} {\bibinfo {author} {\bibfnamefont {L.}~\bibnamefont
  {Zhang}}, \bibinfo {author} {\bibfnamefont {Y.}~\bibnamefont {Yan}}, \bibinfo
  {author} {\bibfnamefont {C.-Q.}\ \bibnamefont {Wu}}, \bibinfo {author}
  {\bibfnamefont {J.-S.}\ \bibnamefont {Wang}}, \ and\ \bibinfo {author}
  {\bibfnamefont {B.}~\bibnamefont {Li}},\ }\href {\doibase
  10.1103/PhysRevB.80.172301} {\bibfield  {journal} {\bibinfo  {journal} {Phys.
  Rev. B}\ }\textbf {\bibinfo {volume} {80}},\ \bibinfo {pages} {172301}
  (\bibinfo {year} {2009})}\BibitemShut {NoStop}%
\bibitem [{\citenamefont {Kuo}\ and\ \citenamefont
  {Chang}(2010)}]{PhysRevB.81.205321}%
  \BibitemOpen
  \bibfield  {author} {\bibinfo {author} {\bibfnamefont {D.~M.-T.}\
  \bibnamefont {Kuo}}\ and\ \bibinfo {author} {\bibfnamefont {Y.-c.}\
  \bibnamefont {Chang}},\ }\href {\doibase 10.1103/PhysRevB.81.205321}
  {\bibfield  {journal} {\bibinfo  {journal} {Phys. Rev. B}\ }\textbf {\bibinfo
  {volume} {81}},\ \bibinfo {pages} {205321} (\bibinfo {year}
  {2010})}\BibitemShut {NoStop}%
\bibitem [{\citenamefont {Otey}\ \emph {et~al.}(2010)\citenamefont {Otey},
  \citenamefont {Lau},\ and\ \citenamefont {Fan}}]{PhysRevLett.104.154301}%
  \BibitemOpen
  \bibfield  {author} {\bibinfo {author} {\bibfnamefont {C.~R.}\ \bibnamefont
  {Otey}}, \bibinfo {author} {\bibfnamefont {W.~T.}\ \bibnamefont {Lau}}, \
  and\ \bibinfo {author} {\bibfnamefont {S.}~\bibnamefont {Fan}},\ }\href
  {\doibase 10.1103/PhysRevLett.104.154301} {\bibfield  {journal} {\bibinfo
  {journal} {Phys. Rev. Lett.}\ }\textbf {\bibinfo {volume} {104}},\ \bibinfo
  {pages} {154301} (\bibinfo {year} {2010})}\BibitemShut {NoStop}%
\bibitem [{\citenamefont {Shen}\ \emph {et~al.}(2011)\citenamefont {Shen},
  \citenamefont {Bradford},\ and\ \citenamefont
  {Shen}}]{PhysRevLett.107.173902}%
  \BibitemOpen
  \bibfield  {author} {\bibinfo {author} {\bibfnamefont {Y.}~\bibnamefont
  {Shen}}, \bibinfo {author} {\bibfnamefont {M.}~\bibnamefont {Bradford}}, \
  and\ \bibinfo {author} {\bibfnamefont {J.-T.}\ \bibnamefont {Shen}},\ }\href
  {\doibase 10.1103/PhysRevLett.107.173902} {\bibfield  {journal} {\bibinfo
  {journal} {Phys. Rev. Lett.}\ }\textbf {\bibinfo {volume} {107}},\ \bibinfo
  {pages} {173902} (\bibinfo {year} {2011})}\BibitemShut {NoStop}%
\bibitem [{\citenamefont {Li}\ \emph {et~al.}(2012)\citenamefont {Li},
  \citenamefont {Ren}, \citenamefont {Wang}, \citenamefont {Zhang},
  \citenamefont {H\"anggi},\ and\ \citenamefont {Li}}]{RevModPhys.84.1045}%
  \BibitemOpen
  \bibfield  {author} {\bibinfo {author} {\bibfnamefont {N.}~\bibnamefont
  {Li}}, \bibinfo {author} {\bibfnamefont {J.}~\bibnamefont {Ren}}, \bibinfo
  {author} {\bibfnamefont {L.}~\bibnamefont {Wang}}, \bibinfo {author}
  {\bibfnamefont {G.}~\bibnamefont {Zhang}}, \bibinfo {author} {\bibfnamefont
  {P.}~\bibnamefont {H\"anggi}}, \ and\ \bibinfo {author} {\bibfnamefont
  {B.}~\bibnamefont {Li}},\ }\href {\doibase 10.1103/RevModPhys.84.1045}
  {\bibfield  {journal} {\bibinfo  {journal} {Rev. Mod. Phys.}\ }\textbf
  {\bibinfo {volume} {84}},\ \bibinfo {pages} {1045} (\bibinfo {year}
  {2012})}\BibitemShut {NoStop}%
\bibitem [{\citenamefont {Ren}\ and\ \citenamefont
  {Zhu}(2013)}]{PhysRevB.88.094427}%
  \BibitemOpen
  \bibfield  {author} {\bibinfo {author} {\bibfnamefont {J.}~\bibnamefont
  {Ren}}\ and\ \bibinfo {author} {\bibfnamefont {J.-X.}\ \bibnamefont {Zhu}},\
  }\href {\doibase 10.1103/PhysRevB.88.094427} {\bibfield  {journal} {\bibinfo
  {journal} {Phys. Rev. B}\ }\textbf {\bibinfo {volume} {88}},\ \bibinfo
  {pages} {094427} (\bibinfo {year} {2013})}\BibitemShut {NoStop}%
\bibitem [{\citenamefont {Tseng}\ \emph {et~al.}(2013)\citenamefont {Tseng},
  \citenamefont {Kuo}, \citenamefont {Chang},\ and\ \citenamefont
  {Lin}}]{doi:10.1063/1.4817258}%
  \BibitemOpen
  \bibfield  {author} {\bibinfo {author} {\bibfnamefont {Y.-C.}\ \bibnamefont
  {Tseng}}, \bibinfo {author} {\bibfnamefont {D.~M.~T.}\ \bibnamefont {Kuo}},
  \bibinfo {author} {\bibfnamefont {Y.-c.}\ \bibnamefont {Chang}}, \ and\
  \bibinfo {author} {\bibfnamefont {Y.-T.}\ \bibnamefont {Lin}},\ }\href
  {\doibase 10.1063/1.4817258} {\bibfield  {journal} {\bibinfo  {journal}
  {Appl. Phys. Lett.}\ }\textbf {\bibinfo {volume} {103}},\ \bibinfo {pages}
  {053108} (\bibinfo {year} {2013})}\BibitemShut {NoStop}%
\bibitem [{\citenamefont {Zhang}\ \emph {et~al.}(2013)\citenamefont {Zhang},
  \citenamefont {Thingna}, \citenamefont {He}, \citenamefont {Wang},\ and\
  \citenamefont {Li}}]{Zhang_2013}%
  \BibitemOpen
  \bibfield  {author} {\bibinfo {author} {\bibfnamefont {L.}~\bibnamefont
  {Zhang}}, \bibinfo {author} {\bibfnamefont {J.}~\bibnamefont {Thingna}},
  \bibinfo {author} {\bibfnamefont {D.}~\bibnamefont {He}}, \bibinfo {author}
  {\bibfnamefont {J.-S.}\ \bibnamefont {Wang}}, \ and\ \bibinfo {author}
  {\bibfnamefont {B.}~\bibnamefont {Li}},\ }\href {\doibase
  10.1209/0295-5075/103/64002} {\bibfield  {journal} {\bibinfo  {journal}
  {EPL}\ }\textbf {\bibinfo {volume} {103}},\ \bibinfo {pages} {64002}
  (\bibinfo {year} {2013})}\BibitemShut {NoStop}%
\bibitem [{\citenamefont {Thingna}\ and\ \citenamefont
  {Wang}(2013)}]{Thingna_2013}%
  \BibitemOpen
  \bibfield  {author} {\bibinfo {author} {\bibfnamefont {J.}~\bibnamefont
  {Thingna}}\ and\ \bibinfo {author} {\bibfnamefont {J.-S.}\ \bibnamefont
  {Wang}},\ }\href {\doibase 10.1209/0295-5075/104/37006} {\bibfield  {journal}
  {\bibinfo  {journal} {EPL}\ }\textbf {\bibinfo {volume} {104}},\ \bibinfo
  {pages} {37006} (\bibinfo {year} {2013})}\BibitemShut {NoStop}%
\bibitem [{\citenamefont {Landi}\ \emph {et~al.}(2014)\citenamefont {Landi},
  \citenamefont {Novais}, \citenamefont {de~Oliveira},\ and\ \citenamefont
  {Karevski}}]{XXZ2}%
  \BibitemOpen
  \bibfield  {author} {\bibinfo {author} {\bibfnamefont {G.~T.}\ \bibnamefont
  {Landi}}, \bibinfo {author} {\bibfnamefont {E.}~\bibnamefont {Novais}},
  \bibinfo {author} {\bibfnamefont {M.~J.}\ \bibnamefont {de~Oliveira}}, \ and\
  \bibinfo {author} {\bibfnamefont {D.}~\bibnamefont {Karevski}},\ }\href
  {\doibase 10.1103/PhysRevE.90.042142} {\bibfield  {journal} {\bibinfo
  {journal} {Phys. Rev. E}\ }\textbf {\bibinfo {volume} {90}},\ \bibinfo
  {pages} {042142} (\bibinfo {year} {2014})}\BibitemShut {NoStop}%
\bibitem [{\citenamefont {Werlang}\ \emph {et~al.}(2014)\citenamefont
  {Werlang}, \citenamefont {Marchiori}, \citenamefont {Cornelio},\ and\
  \citenamefont {Valente}}]{PhysRevE.89.062109}%
  \BibitemOpen
  \bibfield  {author} {\bibinfo {author} {\bibfnamefont {T.}~\bibnamefont
  {Werlang}}, \bibinfo {author} {\bibfnamefont {M.~A.}\ \bibnamefont
  {Marchiori}}, \bibinfo {author} {\bibfnamefont {M.~F.}\ \bibnamefont
  {Cornelio}}, \ and\ \bibinfo {author} {\bibfnamefont {D.}~\bibnamefont
  {Valente}},\ }\href {\doibase 10.1103/PhysRevE.89.062109} {\bibfield
  {journal} {\bibinfo  {journal} {Phys. Rev. E}\ }\textbf {\bibinfo {volume}
  {89}},\ \bibinfo {pages} {062109} (\bibinfo {year} {2014})}\BibitemShut
  {NoStop}%
\bibitem [{\citenamefont {Jiang}\ \emph {et~al.}(2015)\citenamefont {Jiang},
  \citenamefont {Kulkarni}, \citenamefont {Segal},\ and\ \citenamefont
  {Imry}}]{PhysRevB.92.045309}%
  \BibitemOpen
  \bibfield  {author} {\bibinfo {author} {\bibfnamefont {J.-H.}\ \bibnamefont
  {Jiang}}, \bibinfo {author} {\bibfnamefont {M.}~\bibnamefont {Kulkarni}},
  \bibinfo {author} {\bibfnamefont {D.}~\bibnamefont {Segal}}, \ and\ \bibinfo
  {author} {\bibfnamefont {Y.}~\bibnamefont {Imry}},\ }\href {\doibase
  10.1103/PhysRevB.92.045309} {\bibfield  {journal} {\bibinfo  {journal} {Phys.
  Rev. B}\ }\textbf {\bibinfo {volume} {92}},\ \bibinfo {pages} {045309}
  (\bibinfo {year} {2015})}\BibitemShut {NoStop}%
\bibitem [{\citenamefont {Man}\ \emph {et~al.}(2016)\citenamefont {Man},
  \citenamefont {An},\ and\ \citenamefont {Xia}}]{PhysRevE.94.042135}%
  \BibitemOpen
  \bibfield  {author} {\bibinfo {author} {\bibfnamefont {Z.-X.}\ \bibnamefont
  {Man}}, \bibinfo {author} {\bibfnamefont {N.~B.}\ \bibnamefont {An}}, \ and\
  \bibinfo {author} {\bibfnamefont {Y.-J.}\ \bibnamefont {Xia}},\ }\href
  {\doibase 10.1103/PhysRevE.94.042135} {\bibfield  {journal} {\bibinfo
  {journal} {Phys. Rev. E}\ }\textbf {\bibinfo {volume} {94}},\ \bibinfo
  {pages} {042135} (\bibinfo {year} {2016})}\BibitemShut {NoStop}%
\bibitem [{\citenamefont {Schuab}\ \emph {et~al.}(2016)\citenamefont {Schuab},
  \citenamefont {Pereira},\ and\ \citenamefont {Landi}}]{PhysRevE.94.042122}%
  \BibitemOpen
  \bibfield  {author} {\bibinfo {author} {\bibfnamefont {L.}~\bibnamefont
  {Schuab}}, \bibinfo {author} {\bibfnamefont {E.}~\bibnamefont {Pereira}}, \
  and\ \bibinfo {author} {\bibfnamefont {G.~T.}\ \bibnamefont {Landi}},\ }\href
  {\doibase 10.1103/PhysRevE.94.042122} {\bibfield  {journal} {\bibinfo
  {journal} {Phys. Rev. E}\ }\textbf {\bibinfo {volume} {94}},\ \bibinfo
  {pages} {042122} (\bibinfo {year} {2016})}\BibitemShut {NoStop}%
\bibitem [{\citenamefont {Karimi}\ \emph {et~al.}(2017)\citenamefont {Karimi},
  \citenamefont {Pekola}, \citenamefont {Campisi},\ and\ \citenamefont
  {Fazio}}]{Newexp_1}%
  \BibitemOpen
  \bibfield  {author} {\bibinfo {author} {\bibfnamefont {B.}~\bibnamefont
  {Karimi}}, \bibinfo {author} {\bibfnamefont {J.~P.}\ \bibnamefont {Pekola}},
  \bibinfo {author} {\bibfnamefont {M.}~\bibnamefont {Campisi}}, \ and\
  \bibinfo {author} {\bibfnamefont {R.}~\bibnamefont {Fazio}},\ }\href
  {\doibase 10.1088/2058-9565/aa8330} {\bibfield  {journal} {\bibinfo
  {journal} {Quantum Sci. Technol.}\ }\textbf {\bibinfo {volume} {2}},\
  \bibinfo {pages} {044007} (\bibinfo {year} {2017})}\BibitemShut {NoStop}%
\bibitem [{\citenamefont {Ordonez-Miranda}\ \emph {et~al.}(2017)\citenamefont
  {Ordonez-Miranda}, \citenamefont {Ezzahri},\ and\ \citenamefont
  {Joulain}}]{lindblad_equation}%
  \BibitemOpen
  \bibfield  {author} {\bibinfo {author} {\bibfnamefont {J.}~\bibnamefont
  {Ordonez-Miranda}}, \bibinfo {author} {\bibfnamefont {Y.}~\bibnamefont
  {Ezzahri}}, \ and\ \bibinfo {author} {\bibfnamefont {K.}~\bibnamefont
  {Joulain}},\ }\href {\doibase 10.1103/PhysRevE.95.022128} {\bibfield
  {journal} {\bibinfo  {journal} {Phys. Rev. E}\ }\textbf {\bibinfo {volume}
  {95}},\ \bibinfo {pages} {022128} (\bibinfo {year} {2017})}\BibitemShut
  {NoStop}%
\bibitem [{\citenamefont {Pereira}(2017)}]{PhysRevE.96.012114}%
  \BibitemOpen
  \bibfield  {author} {\bibinfo {author} {\bibfnamefont {E.}~\bibnamefont
  {Pereira}},\ }\href {\doibase 10.1103/PhysRevE.96.012114} {\bibfield
  {journal} {\bibinfo  {journal} {Phys. Rev. E}\ }\textbf {\bibinfo {volume}
  {96}},\ \bibinfo {pages} {012114} (\bibinfo {year} {2017})}\BibitemShut
  {NoStop}%
\bibitem [{\citenamefont {Marcos-Vicioso}\ \emph {et~al.}(2018)\citenamefont
  {Marcos-Vicioso}, \citenamefont {L\'opez-Jurado}, \citenamefont
  {Ruiz-Garcia},\ and\ \citenamefont {S\'anchez}}]{PhysRevB.98.035414}%
  \BibitemOpen
  \bibfield  {author} {\bibinfo {author} {\bibfnamefont {A.}~\bibnamefont
  {Marcos-Vicioso}}, \bibinfo {author} {\bibfnamefont {C.}~\bibnamefont
  {L\'opez-Jurado}}, \bibinfo {author} {\bibfnamefont {M.}~\bibnamefont
  {Ruiz-Garcia}}, \ and\ \bibinfo {author} {\bibfnamefont {R.}~\bibnamefont
  {S\'anchez}},\ }\href {\doibase 10.1103/PhysRevB.98.035414} {\bibfield
  {journal} {\bibinfo  {journal} {Phys. Rev. B}\ }\textbf {\bibinfo {volume}
  {98}},\ \bibinfo {pages} {035414} (\bibinfo {year} {2018})}\BibitemShut
  {NoStop}%
\bibitem [{\citenamefont {Balachandran}\ \emph {et~al.}(2018)\citenamefont
  {Balachandran}, \citenamefont {Benenti}, \citenamefont {Pereira},
  \citenamefont {Casati},\ and\ \citenamefont
  {Poletti}}]{PhysRevLett.120.200603}%
  \BibitemOpen
  \bibfield  {author} {\bibinfo {author} {\bibfnamefont {V.}~\bibnamefont
  {Balachandran}}, \bibinfo {author} {\bibfnamefont {G.}~\bibnamefont
  {Benenti}}, \bibinfo {author} {\bibfnamefont {E.}~\bibnamefont {Pereira}},
  \bibinfo {author} {\bibfnamefont {G.}~\bibnamefont {Casati}}, \ and\ \bibinfo
  {author} {\bibfnamefont {D.}~\bibnamefont {Poletti}},\ }\href {\doibase
  10.1103/PhysRevLett.120.200603} {\bibfield  {journal} {\bibinfo  {journal}
  {Phys. Rev. Lett.}\ }\textbf {\bibinfo {volume} {120}},\ \bibinfo {pages}
  {200603} (\bibinfo {year} {2018})}\BibitemShut {NoStop}%
\bibitem [{\citenamefont {Motz}\ \emph {et~al.}(2018)\citenamefont {Motz},
  \citenamefont {Wiedmann}, \citenamefont {Stockburger},\ and\ \citenamefont
  {Ankerhold}}]{Motz_2018}%
  \BibitemOpen
  \bibfield  {author} {\bibinfo {author} {\bibfnamefont {T.}~\bibnamefont
  {Motz}}, \bibinfo {author} {\bibfnamefont {M.}~\bibnamefont {Wiedmann}},
  \bibinfo {author} {\bibfnamefont {J.~T.}\ \bibnamefont {Stockburger}}, \ and\
  \bibinfo {author} {\bibfnamefont {J.}~\bibnamefont {Ankerhold}},\ }\href
  {\doibase 10.1088/1367-2630/aaea90} {\bibfield  {journal} {\bibinfo
  {journal} {New J. Phys.}\ }\textbf {\bibinfo {volume} {20}},\ \bibinfo
  {pages} {113020} (\bibinfo {year} {2018})}\BibitemShut {NoStop}%
\bibitem [{\citenamefont {Kaushik}\ \emph {et~al.}(2018)\citenamefont
  {Kaushik}, \citenamefont {Kaushik},\ and\ \citenamefont {Marathe}}]{lab}%
  \BibitemOpen
  \bibfield  {author} {\bibinfo {author} {\bibfnamefont {S.}~\bibnamefont
  {Kaushik}}, \bibinfo {author} {\bibfnamefont {S.}~\bibnamefont {Kaushik}}, \
  and\ \bibinfo {author} {\bibfnamefont {R.}~\bibnamefont {Marathe}},\ }\href
  {\doibase 10.1140/epjb/e2018-90038-4} {\bibfield  {journal} {\bibinfo
  {journal} {Euro. Phys. J. B}\ }\textbf {\bibinfo {volume} {91}},\ \bibinfo
  {pages} {87} (\bibinfo {year} {2018})}\BibitemShut {NoStop}%
\bibitem [{\citenamefont {Wang}\ \emph {et~al.}(2019)\citenamefont {Wang},
  \citenamefont {Xu}, \citenamefont {Liu},\ and\ \citenamefont {Gao}}]{Newt_3}%
  \BibitemOpen
  \bibfield  {author} {\bibinfo {author} {\bibfnamefont {C.}~\bibnamefont
  {Wang}}, \bibinfo {author} {\bibfnamefont {D.}~\bibnamefont {Xu}}, \bibinfo
  {author} {\bibfnamefont {H.}~\bibnamefont {Liu}}, \ and\ \bibinfo {author}
  {\bibfnamefont {X.}~\bibnamefont {Gao}},\ }\href {\doibase
  10.1103/PhysRevE.99.042102} {\bibfield  {journal} {\bibinfo  {journal} {Phys.
  Rev. E}\ }\textbf {\bibinfo {volume} {99}},\ \bibinfo {pages} {042102}
  (\bibinfo {year} {2019})}\BibitemShut {NoStop}%
\bibitem [{\citenamefont {Balachandran}\ \emph {et~al.}(2019)\citenamefont
  {Balachandran}, \citenamefont {Benenti}, \citenamefont {Pereira},
  \citenamefont {Casati},\ and\ \citenamefont {Poletti}}]{segmentedchain}%
  \BibitemOpen
  \bibfield  {author} {\bibinfo {author} {\bibfnamefont {V.}~\bibnamefont
  {Balachandran}}, \bibinfo {author} {\bibfnamefont {G.}~\bibnamefont
  {Benenti}}, \bibinfo {author} {\bibfnamefont {E.}~\bibnamefont {Pereira}},
  \bibinfo {author} {\bibfnamefont {G.}~\bibnamefont {Casati}}, \ and\ \bibinfo
  {author} {\bibfnamefont {D.}~\bibnamefont {Poletti}},\ }\href {\doibase
  10.1103/PhysRevE.99.032136} {\bibfield  {journal} {\bibinfo  {journal} {Phys.
  Rev. E}\ }\textbf {\bibinfo {volume} {99}},\ \bibinfo {pages} {032136}
  (\bibinfo {year} {2019})}\BibitemShut {NoStop}%
\bibitem [{\citenamefont {Karg\ifmmode \imath \else~\i \fi{}}\ \emph
  {et~al.}(2019)\citenamefont {Karg\ifmmode \imath \else~\i \fi{}},
  \citenamefont {Naseem}, \citenamefont {Opatrn\'y}, \citenamefont
  {M\"ustecapl\ifmmode \imath \else \i \fi{}o\ifmmode~\breve{g}\else
  \u{g}\fi{}lu},\ and\ \citenamefont {Kurizki}}]{Muhammad}%
  \BibitemOpen
  \bibfield  {author} {\bibinfo {author} {\bibfnamefont {C.}~\bibnamefont
  {Karg\ifmmode \imath \else~\i \fi{}}}, \bibinfo {author} {\bibfnamefont
  {M.~T.}\ \bibnamefont {Naseem}}, \bibinfo {author} {\bibfnamefont {T.~c.~v.}\
  \bibnamefont {Opatrn\'y}}, \bibinfo {author} {\bibfnamefont {O.~E.}\
  \bibnamefont {M\"ustecapl\ifmmode \imath \else \i
  \fi{}o\ifmmode~\breve{g}\else \u{g}\fi{}lu}}, \ and\ \bibinfo {author}
  {\bibfnamefont {G.}~\bibnamefont {Kurizki}},\ }\href {\doibase
  10.1103/PhysRevE.99.042121} {\bibfield  {journal} {\bibinfo  {journal} {Phys.
  Rev. E}\ }\textbf {\bibinfo {volume} {99}},\ \bibinfo {pages} {042121}
  (\bibinfo {year} {2019})}\BibitemShut {NoStop}%
\bibitem [{\citenamefont {Lu}\ \emph {et~al.}(2019)\citenamefont {Lu},
  \citenamefont {Wang}, \citenamefont {Ren}, \citenamefont {Kulkarni},\ and\
  \citenamefont {Jiang}}]{PhysRevB.99.035129}%
  \BibitemOpen
  \bibfield  {author} {\bibinfo {author} {\bibfnamefont {J.}~\bibnamefont
  {Lu}}, \bibinfo {author} {\bibfnamefont {R.}~\bibnamefont {Wang}}, \bibinfo
  {author} {\bibfnamefont {J.}~\bibnamefont {Ren}}, \bibinfo {author}
  {\bibfnamefont {M.}~\bibnamefont {Kulkarni}}, \ and\ \bibinfo {author}
  {\bibfnamefont {J.-H.}\ \bibnamefont {Jiang}},\ }\href {\doibase
  10.1103/PhysRevB.99.035129} {\bibfield  {journal} {\bibinfo  {journal} {Phys.
  Rev. B}\ }\textbf {\bibinfo {volume} {99}},\ \bibinfo {pages} {035129}
  (\bibinfo {year} {2019})}\BibitemShut {NoStop}%
\bibitem [{\citenamefont {Riera-Campeny}\ \emph {et~al.}(2019)\citenamefont
  {Riera-Campeny}, \citenamefont {Mehboudi}, \citenamefont {Pons},\ and\
  \citenamefont {Sanpera}}]{PhysRevE.99.032126}%
  \BibitemOpen
  \bibfield  {author} {\bibinfo {author} {\bibfnamefont {A.}~\bibnamefont
  {Riera-Campeny}}, \bibinfo {author} {\bibfnamefont {M.}~\bibnamefont
  {Mehboudi}}, \bibinfo {author} {\bibfnamefont {M.}~\bibnamefont {Pons}}, \
  and\ \bibinfo {author} {\bibfnamefont {A.}~\bibnamefont {Sanpera}},\ }\href
  {\doibase 10.1103/PhysRevE.99.032126} {\bibfield  {journal} {\bibinfo
  {journal} {Phys. Rev. E}\ }\textbf {\bibinfo {volume} {99}},\ \bibinfo
  {pages} {032126} (\bibinfo {year} {2019})}\BibitemShut {NoStop}%
\bibitem [{\citenamefont {Pereira}(2019)}]{PhysRevE.99.032116}%
  \BibitemOpen
  \bibfield  {author} {\bibinfo {author} {\bibfnamefont {E.}~\bibnamefont
  {Pereira}},\ }\href {\doibase 10.1103/PhysRevE.99.032116} {\bibfield
  {journal} {\bibinfo  {journal} {Phys. Rev. E}\ }\textbf {\bibinfo {volume}
  {99}},\ \bibinfo {pages} {032116} (\bibinfo {year} {2019})}\BibitemShut
  {NoStop}%
\bibitem [{\citenamefont {Naseem}\ \emph {et~al.}(2020)\citenamefont {Naseem},
  \citenamefont {Misra}, \citenamefont {M\"ustecaplio\ifmmode~\breve{g}\else
  \u{g}\fi{}lu},\ and\ \citenamefont {Kurizki}}]{PhysRevResearch.2.033285}%
  \BibitemOpen
  \bibfield  {author} {\bibinfo {author} {\bibfnamefont {M.~T.}\ \bibnamefont
  {Naseem}}, \bibinfo {author} {\bibfnamefont {A.}~\bibnamefont {Misra}},
  \bibinfo {author} {\bibfnamefont {O.~E.}\ \bibnamefont
  {M\"ustecaplio\ifmmode~\breve{g}\else \u{g}\fi{}lu}}, \ and\ \bibinfo
  {author} {\bibfnamefont {G.}~\bibnamefont {Kurizki}},\ }\href {\doibase
  10.1103/PhysRevResearch.2.033285} {\bibfield  {journal} {\bibinfo  {journal}
  {Phys. Rev. Research}\ }\textbf {\bibinfo {volume} {2}},\ \bibinfo {pages}
  {033285} (\bibinfo {year} {2020})}\BibitemShut {NoStop}%
\bibitem [{\citenamefont {Silva}\ \emph {et~al.}(2020)\citenamefont {Silva},
  \citenamefont {Landi}, \citenamefont {Drumond},\ and\ \citenamefont
  {Pereira}}]{PhysRevE.102.062146}%
  \BibitemOpen
  \bibfield  {author} {\bibinfo {author} {\bibfnamefont {S.~H.~S.}\
  \bibnamefont {Silva}}, \bibinfo {author} {\bibfnamefont {G.~T.}\ \bibnamefont
  {Landi}}, \bibinfo {author} {\bibfnamefont {R.~C.}\ \bibnamefont {Drumond}},
  \ and\ \bibinfo {author} {\bibfnamefont {E.}~\bibnamefont {Pereira}},\ }\href
  {\doibase 10.1103/PhysRevE.102.062146} {\bibfield  {journal} {\bibinfo
  {journal} {Phys. Rev. E}\ }\textbf {\bibinfo {volume} {102}},\ \bibinfo
  {pages} {062146} (\bibinfo {year} {2020})}\BibitemShut {NoStop}%
\bibitem [{\citenamefont {Alexander}(2020)}]{PhysRevE.101.062122}%
  \BibitemOpen
  \bibfield  {author} {\bibinfo {author} {\bibfnamefont {T.~J.}\ \bibnamefont
  {Alexander}},\ }\href {\doibase 10.1103/PhysRevE.101.062122} {\bibfield
  {journal} {\bibinfo  {journal} {Phys. Rev. E}\ }\textbf {\bibinfo {volume}
  {101}},\ \bibinfo {pages} {062122} (\bibinfo {year} {2020})}\BibitemShut
  {NoStop}%
\bibitem [{\citenamefont {Xu}\ \emph {et~al.}(2021)\citenamefont {Xu},
  \citenamefont {Stockburger},\ and\ \citenamefont
  {Ankerhold}}]{PhysRevB.103.104304}%
  \BibitemOpen
  \bibfield  {author} {\bibinfo {author} {\bibfnamefont {M.}~\bibnamefont
  {Xu}}, \bibinfo {author} {\bibfnamefont {J.~T.}\ \bibnamefont {Stockburger}},
  \ and\ \bibinfo {author} {\bibfnamefont {J.}~\bibnamefont {Ankerhold}},\
  }\href {\doibase 10.1103/PhysRevB.103.104304} {\bibfield  {journal} {\bibinfo
   {journal} {Phys. Rev. B}\ }\textbf {\bibinfo {volume} {103}},\ \bibinfo
  {pages} {104304} (\bibinfo {year} {2021})}\BibitemShut {NoStop}%
\bibitem [{\citenamefont {Kalantar}\ \emph {et~al.}(2021)\citenamefont
  {Kalantar}, \citenamefont {Agarwalla},\ and\ \citenamefont
  {Segal}}]{PhysRevE.103.052130}%
  \BibitemOpen
  \bibfield  {author} {\bibinfo {author} {\bibfnamefont {N.}~\bibnamefont
  {Kalantar}}, \bibinfo {author} {\bibfnamefont {B.~K.}\ \bibnamefont
  {Agarwalla}}, \ and\ \bibinfo {author} {\bibfnamefont {D.}~\bibnamefont
  {Segal}},\ }\href {\doibase 10.1103/PhysRevE.103.052130} {\bibfield
  {journal} {\bibinfo  {journal} {Phys. Rev. E}\ }\textbf {\bibinfo {volume}
  {103}},\ \bibinfo {pages} {052130} (\bibinfo {year} {2021})}\BibitemShut
  {NoStop}%
\bibitem [{\citenamefont {Bhandari}\ \emph {et~al.}(2021)\citenamefont
  {Bhandari}, \citenamefont {Erdman}, \citenamefont {Fazio}, \citenamefont
  {Paladino},\ and\ \citenamefont {Taddei}}]{PhysRevB.103.155434}%
  \BibitemOpen
  \bibfield  {author} {\bibinfo {author} {\bibfnamefont {B.}~\bibnamefont
  {Bhandari}}, \bibinfo {author} {\bibfnamefont {P.~A.}\ \bibnamefont
  {Erdman}}, \bibinfo {author} {\bibfnamefont {R.}~\bibnamefont {Fazio}},
  \bibinfo {author} {\bibfnamefont {E.}~\bibnamefont {Paladino}}, \ and\
  \bibinfo {author} {\bibfnamefont {F.}~\bibnamefont {Taddei}},\ }\href
  {\doibase 10.1103/PhysRevB.103.155434} {\bibfield  {journal} {\bibinfo
  {journal} {Phys. Rev. B}\ }\textbf {\bibinfo {volume} {103}},\ \bibinfo
  {pages} {155434} (\bibinfo {year} {2021})}\BibitemShut {NoStop}%
\bibitem [{\citenamefont {Iorio}\ \emph {et~al.}(2021)\citenamefont {Iorio},
  \citenamefont {Strambini}, \citenamefont {Haack}, \citenamefont {Campisi},\
  and\ \citenamefont {Giazotto}}]{PhysRevApplied.15.054050}%
  \BibitemOpen
  \bibfield  {author} {\bibinfo {author} {\bibfnamefont {A.}~\bibnamefont
  {Iorio}}, \bibinfo {author} {\bibfnamefont {E.}~\bibnamefont {Strambini}},
  \bibinfo {author} {\bibfnamefont {G.}~\bibnamefont {Haack}}, \bibinfo
  {author} {\bibfnamefont {M.}~\bibnamefont {Campisi}}, \ and\ \bibinfo
  {author} {\bibfnamefont {F.}~\bibnamefont {Giazotto}},\ }\href {\doibase
  10.1103/PhysRevApplied.15.054050} {\bibfield  {journal} {\bibinfo  {journal}
  {Phys. Rev. Applied}\ }\textbf {\bibinfo {volume} {15}},\ \bibinfo {pages}
  {054050} (\bibinfo {year} {2021})}\BibitemShut {NoStop}%
\bibitem [{\citenamefont {Sim\'on}\ \emph {et~al.}(2021)\citenamefont
  {Sim\'on}, \citenamefont {Ala\~na}, \citenamefont {Pons}, \citenamefont
  {Ruiz-Garc\'{\i}a},\ and\ \citenamefont {Muga}}]{PhysRevE.103.012134}%
  \BibitemOpen
  \bibfield  {author} {\bibinfo {author} {\bibfnamefont {M.~A.}\ \bibnamefont
  {Sim\'on}}, \bibinfo {author} {\bibfnamefont {A.}~\bibnamefont {Ala\~na}},
  \bibinfo {author} {\bibfnamefont {M.}~\bibnamefont {Pons}}, \bibinfo {author}
  {\bibfnamefont {A.}~\bibnamefont {Ruiz-Garc\'{\i}a}}, \ and\ \bibinfo
  {author} {\bibfnamefont {J.~G.}\ \bibnamefont {Muga}},\ }\href {\doibase
  10.1103/PhysRevE.103.012134} {\bibfield  {journal} {\bibinfo  {journal}
  {Phys. Rev. E}\ }\textbf {\bibinfo {volume} {103}},\ \bibinfo {pages}
  {012134} (\bibinfo {year} {2021})}\BibitemShut {NoStop}%
\bibitem [{\citenamefont {Stevenson}\ and\ \citenamefont
  {Braunecker}(2021)}]{PhysRevB.103.115413}%
  \BibitemOpen
  \bibfield  {author} {\bibinfo {author} {\bibfnamefont {C.}~\bibnamefont
  {Stevenson}}\ and\ \bibinfo {author} {\bibfnamefont {B.}~\bibnamefont
  {Braunecker}},\ }\href {\doibase 10.1103/PhysRevB.103.115413} {\bibfield
  {journal} {\bibinfo  {journal} {Phys. Rev. B}\ }\textbf {\bibinfo {volume}
  {103}},\ \bibinfo {pages} {115413} (\bibinfo {year} {2021})}\BibitemShut
  {NoStop}%
\bibitem [{\citenamefont {Tupkary}\ \emph {et~al.}(2021)\citenamefont
  {Tupkary}, \citenamefont {Dhar}, \citenamefont {Kulkarni},\ and\
  \citenamefont {Purkayastha}}]{tupkary2021}%
  \BibitemOpen
  \bibfield  {author} {\bibinfo {author} {\bibfnamefont {D.}~\bibnamefont
  {Tupkary}}, \bibinfo {author} {\bibfnamefont {A.}~\bibnamefont {Dhar}},
  \bibinfo {author} {\bibfnamefont {M.}~\bibnamefont {Kulkarni}}, \ and\
  \bibinfo {author} {\bibfnamefont {A.}~\bibnamefont {Purkayastha}},\
  }\href@noop {} {\enquote {\bibinfo {title} {Fundamental limitations in
  lindblad descriptions of systems weakly coupled to baths},}\ } (\bibinfo
  {year} {2021}),\ \Eprint {http://arxiv.org/abs/2105.12091} {arXiv:2105.12091
  [quant-ph]} \BibitemShut {NoStop}%
\bibitem [{\citenamefont {Yang}\ \emph {et~al.}(2009)\citenamefont {Yang},
  \citenamefont {Zhang},\ and\ \citenamefont {Li}}]{experiment4}%
  \BibitemOpen
  \bibfield  {author} {\bibinfo {author} {\bibfnamefont {N.}~\bibnamefont
  {Yang}}, \bibinfo {author} {\bibfnamefont {G.}~\bibnamefont {Zhang}}, \ and\
  \bibinfo {author} {\bibfnamefont {B.}~\bibnamefont {Li}},\ }\href {\doibase
  10.1063/1.3183587} {\bibfield  {journal} {\bibinfo  {journal} {Appl. Phys.
  Lett.}\ }\textbf {\bibinfo {volume} {95}},\ \bibinfo {pages} {033107}
  (\bibinfo {year} {2009})}\BibitemShut {NoStop}%
\bibitem [{\citenamefont {Jiang}\ \emph {et~al.}(2010)\citenamefont {Jiang},
  \citenamefont {Wang},\ and\ \citenamefont {Li}}]{JWJing_experiment}%
  \BibitemOpen
  \bibfield  {author} {\bibinfo {author} {\bibfnamefont {J.~W.}\ \bibnamefont
  {Jiang}}, \bibinfo {author} {\bibfnamefont {J.~S.}\ \bibnamefont {Wang}}, \
  and\ \bibinfo {author} {\bibfnamefont {B.}~\bibnamefont {Li}},\ }\href
  {\doibase 10.1209/0295-5075/89/46005} {\bibfield  {journal} {\bibinfo
  {journal} {Europhys. Lett.}\ }\textbf {\bibinfo {volume} {89}},\ \bibinfo
  {pages} {46005} (\bibinfo {year} {2010})}\BibitemShut {NoStop}%
\bibitem [{\citenamefont {Chen}\ \emph {et~al.}(2014)\citenamefont {Chen},
  \citenamefont {Wong}, \citenamefont {Lubner}, \citenamefont {Yee},
  \citenamefont {Miller}, \citenamefont {Jang}, \citenamefont {Hardin},
  \citenamefont {Fong}, \citenamefont {Garay},\ and\ \citenamefont
  {Dames}}]{Chen2014}%
  \BibitemOpen
  \bibfield  {author} {\bibinfo {author} {\bibfnamefont {Z.}~\bibnamefont
  {Chen}}, \bibinfo {author} {\bibfnamefont {C.}~\bibnamefont {Wong}}, \bibinfo
  {author} {\bibfnamefont {S.}~\bibnamefont {Lubner}}, \bibinfo {author}
  {\bibfnamefont {S.}~\bibnamefont {Yee}}, \bibinfo {author} {\bibfnamefont
  {J.}~\bibnamefont {Miller}}, \bibinfo {author} {\bibfnamefont
  {W.}~\bibnamefont {Jang}}, \bibinfo {author} {\bibfnamefont {C.}~\bibnamefont
  {Hardin}}, \bibinfo {author} {\bibfnamefont {A.}~\bibnamefont {Fong}},
  \bibinfo {author} {\bibfnamefont {J.~E.}\ \bibnamefont {Garay}}, \ and\
  \bibinfo {author} {\bibfnamefont {C.}~\bibnamefont {Dames}},\ }\href
  {\doibase 10.1038/ncomms6446} {\bibfield  {journal} {\bibinfo  {journal}
  {Nat. Commun.}\ }\textbf {\bibinfo {volume} {5}},\ \bibinfo {pages} {5446}
  (\bibinfo {year} {2014})}\BibitemShut {NoStop}%
\bibitem [{\citenamefont {Maria José Martínez-Pérez}(2015)}]{Martinez15}%
  \BibitemOpen
  \bibfield  {author} {\bibinfo {author} {\bibfnamefont {F.~G.}\ \bibnamefont
  {Maria José Martínez-Pérez}, \bibfnamefont {Antonio~Fornieri}},\ }\href
  {\doibase 10.1038/nnano.2015.11} {\bibfield  {journal} {\bibinfo  {journal}
  {Nature Nanotechnology}\ }\textbf {\bibinfo {volume} {10}},\ \bibinfo {pages}
  {1748} (\bibinfo {year} {2015})}\BibitemShut {NoStop}%
\bibitem [{\citenamefont {Seif}\ \emph {et~al.}(2018)\citenamefont {Seif},
  \citenamefont {DeGottardi}, \citenamefont {Esfarjani},\ and\ \citenamefont
  {Hafezi}}]{Seif2018}%
  \BibitemOpen
  \bibfield  {author} {\bibinfo {author} {\bibfnamefont {A.}~\bibnamefont
  {Seif}}, \bibinfo {author} {\bibfnamefont {W.}~\bibnamefont {DeGottardi}},
  \bibinfo {author} {\bibfnamefont {K.}~\bibnamefont {Esfarjani}}, \ and\
  \bibinfo {author} {\bibfnamefont {M.}~\bibnamefont {Hafezi}},\ }\href
  {\doibase 10.1038/s41467-018-03624-y} {\bibfield  {journal} {\bibinfo
  {journal} {Nat. Commun.}\ }\textbf {\bibinfo {volume} {9}},\ \bibinfo {pages}
  {1207} (\bibinfo {year} {2018})}\BibitemShut {NoStop}%
\bibitem [{\citenamefont {Ronzani}\ \emph {et~al.}(2018)\citenamefont
  {Ronzani}, \citenamefont {Karimi}, \citenamefont {Senior}, \citenamefont
  {Chang}, \citenamefont {Peltonen}, \citenamefont {Chen},\ and\ \citenamefont
  {Pekola}}]{Newexp_3}%
  \BibitemOpen
  \bibfield  {author} {\bibinfo {author} {\bibfnamefont {A.}~\bibnamefont
  {Ronzani}}, \bibinfo {author} {\bibfnamefont {B.}~\bibnamefont {Karimi}},
  \bibinfo {author} {\bibfnamefont {J.}~\bibnamefont {Senior}}, \bibinfo
  {author} {\bibfnamefont {Y.-C.}\ \bibnamefont {Chang}}, \bibinfo {author}
  {\bibfnamefont {J.~T.}\ \bibnamefont {Peltonen}}, \bibinfo {author}
  {\bibfnamefont {C.}~\bibnamefont {Chen}}, \ and\ \bibinfo {author}
  {\bibfnamefont {J.~P.}\ \bibnamefont {Pekola}},\ }\href {\doibase
  10.1038/s41567-018-0199-4} {\bibfield  {journal} {\bibinfo  {journal} {Nat.
  Phys.}\ }\textbf {\bibinfo {volume} {14}},\ \bibinfo {pages} {991} (\bibinfo
  {year} {2018})}\BibitemShut {NoStop}%
\bibitem [{\citenamefont {Senior}\ \emph {et~al.}(2020)\citenamefont {Senior},
  \citenamefont {Gubaydullin}, \citenamefont {Karimi}, \citenamefont
  {Peltonen}, \citenamefont {Ankerhold},\ and\ \citenamefont
  {Pekola}}]{Senior2020}%
  \BibitemOpen
  \bibfield  {author} {\bibinfo {author} {\bibfnamefont {J.}~\bibnamefont
  {Senior}}, \bibinfo {author} {\bibfnamefont {A.}~\bibnamefont {Gubaydullin}},
  \bibinfo {author} {\bibfnamefont {B.}~\bibnamefont {Karimi}}, \bibinfo
  {author} {\bibfnamefont {J.~T.}\ \bibnamefont {Peltonen}}, \bibinfo {author}
  {\bibfnamefont {J.}~\bibnamefont {Ankerhold}}, \ and\ \bibinfo {author}
  {\bibfnamefont {J.~P.}\ \bibnamefont {Pekola}},\ }\href {\doibase
  10.1038/s42005-020-0307-5} {\bibfield  {journal} {\bibinfo  {journal}
  {Commun. Phys.}\ }\textbf {\bibinfo {volume} {3}},\ \bibinfo {pages} {40}
  (\bibinfo {year} {2020})}\BibitemShut {NoStop}%
\bibitem [{\citenamefont {Maillet}\ \emph {et~al.}(2020)\citenamefont
  {Maillet}, \citenamefont {Subero}, \citenamefont {Peltonen}, \citenamefont
  {Golubev},\ and\ \citenamefont {Pekola}}]{Maillet2020}%
  \BibitemOpen
  \bibfield  {author} {\bibinfo {author} {\bibfnamefont {O.}~\bibnamefont
  {Maillet}}, \bibinfo {author} {\bibfnamefont {D.}~\bibnamefont {Subero}},
  \bibinfo {author} {\bibfnamefont {J.~T.}\ \bibnamefont {Peltonen}}, \bibinfo
  {author} {\bibfnamefont {D.~S.}\ \bibnamefont {Golubev}}, \ and\ \bibinfo
  {author} {\bibfnamefont {J.~P.}\ \bibnamefont {Pekola}},\ }\href {\doibase
  10.1038/s41467-020-18163-8} {\bibfield  {journal} {\bibinfo  {journal} {Nat.
  Commun.}\ }\textbf {\bibinfo {volume} {11}},\ \bibinfo {pages} {4326}
  (\bibinfo {year} {2020})}\BibitemShut {NoStop}%
\bibitem [{\citenamefont {Dhar}(2008)}]{dhar}%
  \BibitemOpen
  \bibfield  {author} {\bibinfo {author} {\bibfnamefont {A.}~\bibnamefont
  {Dhar}},\ }\href {\doibase 10.1080/00018730802538522} {\bibfield  {journal}
  {\bibinfo  {journal} {Adv. Phys.}\ }\textbf {\bibinfo {volume} {57}},\
  \bibinfo {pages} {457} (\bibinfo {year} {2008})}\BibitemShut {NoStop}%
\bibitem [{\citenamefont
  {Dzyaloshinsky}(1958)}]{dzyaloshinsky_thermodynamic_1958}%
  \BibitemOpen
  \bibfield  {author} {\bibinfo {author} {\bibfnamefont {I.}~\bibnamefont
  {Dzyaloshinsky}},\ }\href {\doibase 10.1016/0022-3697(58)90076-3} {\bibfield
  {journal} {\bibinfo  {journal} {J. Phys. Chem. Solids}\ }\textbf {\bibinfo
  {volume} {4}},\ \bibinfo {pages} {241} (\bibinfo {year} {1958})}\BibitemShut
  {NoStop}%
\bibitem [{\citenamefont {Moriya}(1960)}]{moriya_anisotropic_1960}%
  \BibitemOpen
  \bibfield  {author} {\bibinfo {author} {\bibfnamefont {T.}~\bibnamefont
  {Moriya}},\ }\href {\doibase 10.1103/PhysRev.120.91} {\bibfield  {journal}
  {\bibinfo  {journal} {Phys. Rev.}\ }\textbf {\bibinfo {volume} {120}},\
  \bibinfo {pages} {91} (\bibinfo {year} {1960})}\BibitemShut {NoStop}%
\bibitem [{\citenamefont {Hui-Ping}\ \emph {et~al.}(2006)\citenamefont
  {Hui-Ping}, \citenamefont {Yun-Zhou},\ and\ \citenamefont
  {Lin}}]{Hui_Ping_2006}%
  \BibitemOpen
  \bibfield  {author} {\bibinfo {author} {\bibfnamefont {L.}~\bibnamefont
  {Hui-Ping}}, \bibinfo {author} {\bibfnamefont {S.}~\bibnamefont {Yun-Zhou}},
  \ and\ \bibinfo {author} {\bibfnamefont {Y.}~\bibnamefont {Lin}},\ }\href
  {\doibase 10.1088/0256-307x/23/7/016} {\bibfield  {journal} {\bibinfo
  {journal} {Chin. Phys. Lett.}\ }\textbf {\bibinfo {volume} {23}},\ \bibinfo
  {pages} {1713} (\bibinfo {year} {2006})}\BibitemShut {NoStop}%
\bibitem [{\citenamefont {Li}\ and\ \citenamefont {Tong}(2012)}]{Li2012}%
  \BibitemOpen
  \bibfield  {author} {\bibinfo {author} {\bibfnamefont {Z.}~\bibnamefont {Li},
  \bibfnamefont {W.and~Zhang}}\ and\ \bibinfo {author} {\bibfnamefont
  {P.}~\bibnamefont {Tong}},\ }\href {\doibase 10.1140/epjb/e2012-20798-6}
  {\bibfield  {journal} {\bibinfo  {journal} {Euro. Phys. J. B}\ }\textbf
  {\bibinfo {volume} {85}},\ \bibinfo {pages} {73} (\bibinfo {year}
  {2012})}\BibitemShut {NoStop}%
\bibitem [{\citenamefont {Chen}\ and\ \citenamefont {Wang}(2015)}]{CHEN201558}%
  \BibitemOpen
  \bibfield  {author} {\bibinfo {author} {\bibfnamefont {T.}~\bibnamefont
  {Chen}}\ and\ \bibinfo {author} {\bibfnamefont {X.-B.}\ \bibnamefont
  {Wang}},\ }\href {\doibase https://doi.org/10.1016/j.physe.2015.04.021}
  {\bibfield  {journal} {\bibinfo  {journal} {Physica E}\ }\textbf {\bibinfo
  {volume} {72}},\ \bibinfo {pages} {58} (\bibinfo {year} {2015})}\BibitemShut
  {NoStop}%
\bibitem [{\citenamefont {Xu}\ \emph {et~al.}(2016)\citenamefont {Xu},
  \citenamefont {Liu},\ and\ \citenamefont {Luo}}]{XU2016107}%
  \BibitemOpen
  \bibfield  {author} {\bibinfo {author} {\bibfnamefont {A.-H.}\ \bibnamefont
  {Xu}}, \bibinfo {author} {\bibfnamefont {J.}~\bibnamefont {Liu}}, \ and\
  \bibinfo {author} {\bibfnamefont {B.}~\bibnamefont {Luo}},\ }\href {\doibase
  https://doi.org/10.1016/j.physb.2016.07.007} {\bibfield  {journal} {\bibinfo
  {journal} {Physica B}\ }\textbf {\bibinfo {volume} {499}},\ \bibinfo {pages}
  {107} (\bibinfo {year} {2016})}\BibitemShut {NoStop}%
\bibitem [{\citenamefont {Levy}\ and\ \citenamefont
  {Kosloff}(2014)}]{Levy_2014}%
  \BibitemOpen
  \bibfield  {author} {\bibinfo {author} {\bibfnamefont {A.}~\bibnamefont
  {Levy}}\ and\ \bibinfo {author} {\bibfnamefont {R.}~\bibnamefont {Kosloff}},\
  }\href {\doibase 10.1209/0295-5075/107/20004} {\bibfield  {journal} {\bibinfo
   {journal} {Europhys. Lett.}\ }\textbf {\bibinfo {volume} {107}},\ \bibinfo
  {pages} {20004} (\bibinfo {year} {2014})}\BibitemShut {NoStop}%
\bibitem [{\citenamefont {Naseem}\ \emph {et~al.}(2018)\citenamefont {Naseem},
  \citenamefont {Xuereb},\ and\ \citenamefont {M\"ustecapl\ifmmode \imath \else
  \i \fi{}o\ifmmode~\breve{g}\else \u{g}\fi{}lu}}]{PhysRevA.98.052123}%
  \BibitemOpen
  \bibfield  {author} {\bibinfo {author} {\bibfnamefont {M.~T.}\ \bibnamefont
  {Naseem}}, \bibinfo {author} {\bibfnamefont {A.}~\bibnamefont {Xuereb}}, \
  and\ \bibinfo {author} {\bibfnamefont {O.~E.}\ \bibnamefont
  {M\"ustecapl\ifmmode \imath \else \i \fi{}o\ifmmode~\breve{g}\else
  \u{g}\fi{}lu}},\ }\href {\doibase 10.1103/PhysRevA.98.052123} {\bibfield
  {journal} {\bibinfo  {journal} {Phys. Rev. A}\ }\textbf {\bibinfo {volume}
  {98}},\ \bibinfo {pages} {052123} (\bibinfo {year} {2018})}\BibitemShut
  {NoStop}%
\bibitem [{\citenamefont {Micadei}\ \emph {et~al.}(2019)\citenamefont
  {Micadei}, \citenamefont {Peterson}, \citenamefont {Souza}, \citenamefont
  {Sarthour}, \citenamefont {Oliveira}, \citenamefont {Landi}, \citenamefont
  {Batalh{\~a}o}, \citenamefont {Serra},\ and\ \citenamefont
  {Lutz}}]{NatureDMexperiment}%
  \BibitemOpen
  \bibfield  {author} {\bibinfo {author} {\bibfnamefont {K.}~\bibnamefont
  {Micadei}}, \bibinfo {author} {\bibfnamefont {J.~P.~S.}\ \bibnamefont
  {Peterson}}, \bibinfo {author} {\bibfnamefont {A.~M.}\ \bibnamefont {Souza}},
  \bibinfo {author} {\bibfnamefont {R.~S.}\ \bibnamefont {Sarthour}}, \bibinfo
  {author} {\bibfnamefont {I.~S.}\ \bibnamefont {Oliveira}}, \bibinfo {author}
  {\bibfnamefont {G.~T.}\ \bibnamefont {Landi}}, \bibinfo {author}
  {\bibfnamefont {T.~B.}\ \bibnamefont {Batalh{\~a}o}}, \bibinfo {author}
  {\bibfnamefont {R.~M.}\ \bibnamefont {Serra}}, \ and\ \bibinfo {author}
  {\bibfnamefont {E.}~\bibnamefont {Lutz}},\ }\href {\doibase
  10.1038/s41467-019-10333-7} {\bibfield  {journal} {\bibinfo  {journal} {Nat.
  Commun.}\ }\textbf {\bibinfo {volume} {10}},\ \bibinfo {pages} {2456}
  (\bibinfo {year} {2019})}\BibitemShut {NoStop}%
\bibitem [{\citenamefont {Breuer}\ and\ \citenamefont
  {Petruccione}(2002)}]{breuer2002}%
  \BibitemOpen
  \bibfield  {author} {\bibinfo {author} {\bibfnamefont {H.~P.}\ \bibnamefont
  {Breuer}}\ and\ \bibinfo {author} {\bibfnamefont {F.}~\bibnamefont
  {Petruccione}},\ }\href@noop {} {\emph {\bibinfo {title} {The theory of open
  quantum systems}}}\ (\bibinfo  {publisher} {Oxford university press},\
  \bibinfo {year} {2002})\BibitemShut {NoStop}%
\bibitem [{\citenamefont {Levy}\ \emph {et~al.}(2012)\citenamefont {Levy},
  \citenamefont {Alicki},\ and\ \citenamefont {Kosloff}}]{PhysRevE.85.061126}%
  \BibitemOpen
  \bibfield  {author} {\bibinfo {author} {\bibfnamefont {A.}~\bibnamefont
  {Levy}}, \bibinfo {author} {\bibfnamefont {R.}~\bibnamefont {Alicki}}, \ and\
  \bibinfo {author} {\bibfnamefont {R.}~\bibnamefont {Kosloff}},\ }\href
  {\doibase 10.1103/PhysRevE.85.061126} {\bibfield  {journal} {\bibinfo
  {journal} {Phys. Rev. E}\ }\textbf {\bibinfo {volume} {85}},\ \bibinfo
  {pages} {061126} (\bibinfo {year} {2012})}\BibitemShut {NoStop}%
\bibitem [{\citenamefont {Cattaneo}\ \emph {et~al.}(2019)\citenamefont
  {Cattaneo}, \citenamefont {Giorgi}, \citenamefont {Maniscalco},\ and\
  \citenamefont {Zambrini}}]{Cattaneo_2019}%
  \BibitemOpen
  \bibfield  {author} {\bibinfo {author} {\bibfnamefont {M.}~\bibnamefont
  {Cattaneo}}, \bibinfo {author} {\bibfnamefont {G.~L.}\ \bibnamefont
  {Giorgi}}, \bibinfo {author} {\bibfnamefont {S.}~\bibnamefont {Maniscalco}},
  \ and\ \bibinfo {author} {\bibfnamefont {R.}~\bibnamefont {Zambrini}},\
  }\href {\doibase 10.1088/1367-2630/ab54ac} {\bibfield  {journal} {\bibinfo
  {journal} {New J. Phys.}\ }\textbf {\bibinfo {volume} {21}},\ \bibinfo
  {pages} {113045} (\bibinfo {year} {2019})}\BibitemShut {NoStop}%
\bibitem [{\citenamefont {Kosloff}(2013)}]{e15062100}%
  \BibitemOpen
  \bibfield  {author} {\bibinfo {author} {\bibfnamefont {R.}~\bibnamefont
  {Kosloff}},\ }\href {\doibase 10.3390/e15062100} {\bibfield  {journal}
  {\bibinfo  {journal} {Entropy}\ }\textbf {\bibinfo {volume} {15}},\ \bibinfo
  {pages} {2100} (\bibinfo {year} {2013})}\BibitemShut {NoStop}%
\bibitem [{\citenamefont {Peres}(1996)}]{PhysRevLett.77.1413}%
  \BibitemOpen
  \bibfield  {author} {\bibinfo {author} {\bibfnamefont {A.}~\bibnamefont
  {Peres}},\ }\href {\doibase 10.1103/PhysRevLett.77.1413} {\bibfield
  {journal} {\bibinfo  {journal} {Phys. Rev. Lett.}\ }\textbf {\bibinfo
  {volume} {77}},\ \bibinfo {pages} {1413} (\bibinfo {year}
  {1996})}\BibitemShut {NoStop}%
\bibitem [{\citenamefont {Wootters}(1998)}]{Wootters}%
  \BibitemOpen
  \bibfield  {author} {\bibinfo {author} {\bibfnamefont {W.~K.}\ \bibnamefont
  {Wootters}},\ }\href {\doibase 10.1103/PhysRevLett.80.2245} {\bibfield
  {journal} {\bibinfo  {journal} {Phys. Rev. Lett.}\ }\textbf {\bibinfo
  {volume} {80}},\ \bibinfo {pages} {2245} (\bibinfo {year}
  {1998})}\BibitemShut {NoStop}%
\end{thebibliography}%
	
\end{document}